\documentclass[journal,twoside]{IEEEtranTCOM}
\usepackage{cite}
\usepackage{citesort}
\usepackage{graphicx}
\usepackage{epstopdf}
\usepackage[cmex10]{amsmath}\interdisplaylinepenalty=2500
\usepackage[caption=false]{subfig}
\usepackage{relsize}
\usepackage{nicefrac}
\usepackage{amssymb,bm,upgreek}
\usepackage{algorithm}
\usepackage{algpseudocode}
\usepackage{multicol}
\usepackage{xcolor}

\newtheorem{lemma}{Lemma}
\newtheorem{proposition}{Proposition}
\newtheorem{corollary}{Corollary}
\newtheorem{remark}{Remark}

\begin{document}

\title{Multi-Cell Multiuser Massive MIMO Networks: User Capacity Analysis and Pilot Design}

\author{Noman~Akbar,~\IEEEmembership{Student Member,~IEEE,}
        Nan~Yang,~\IEEEmembership{Member,~IEEE,}\\
        Parastoo Sadeghi,~\IEEEmembership{Senior Member,~IEEE,}
        and Rodney A. Kennedy,~\IEEEmembership{Fellow,~IEEE}%
\thanks{Manuscript received March 16, 2016; revised August 14, 2016; accepted September 22, 2016. The associate editor coordinating the review of this paper and approving it for publication was Y.~J.~Zhang.}
\thanks{This work was presented in part at the IEEE Global Communications Conference (Globecom), Washington, DC, USA, Dec. 2016.}
\thanks{N. Akbar, N. Yang, P. Sadeghi, and R. A. Kennedy are with the Research School of Engineering, Australian National University, Acton, ACT 2601, Australia (e-mail: \{noman.akbar, nan.yang, parastoo.sadeghi, rodney.kennedy\}@anu.edu.au).}
\thanks{Color versions of one or more of the figures in this paper are available online at http://ieeexplore.ieee.org.}}

\markboth{IEEE TRANSACTIONS ON COMMUNICATIONS}{Akbar \MakeLowercase{\textit{et al.}}: Multi-Cell Multiuser Massive MIMO Networks: User Capacity Analysis and Pilot Design}

\maketitle

\begin{abstract}
We propose a novel pilot sequence design to mitigate pilot contamination in multi-cell multiuser massive multiple-input multiple-output networks. Our proposed design generates pilot sequences in the multi-cell network and devises power allocation at base stations (BSs) for downlink transmission. The pilot sequences together with the power allocation ensure that the user capacity of the network is achieved and the pre-defined signal-to-interference-plus-noise ratio (SINR) requirements of all users are met. To realize our design, we first derive new closed-form expressions for the user capacity and the user capacity region. Built upon these expressions, we then develop a new algorithm to obtain the required pilot sequences and power allocation. We further determine the minimum number of antennas required at BSs to achieve certain SINR requirements of all users. Numerical results are presented to corroborate our analysis and to examine the impact of key parameters, such as the pilot sequence length and the total number of users, on the network performance. A pivotal conclusion is reached that our design achieves a larger user capacity region than the existing designs and needs less antennas at the BS to fulfill the pre-defined SINR requirements of all users in the network than the existing designs.
\end{abstract}


\section{Introduction}

\IEEEPARstart{M}{assive} multiple-input multiple-output (MIMO) has been identified as one of the indispensable technologies to support ultra-high data rate for a huge number of mobile users in  the fifth generation wireless systems \cite{Marzetta2010,Boccardi2014,Yang2015}. The key distinguishing feature of massive MIMO from conventional MIMO lies in the very large number of antennas deployed at base stations (BSs), which offers favorable propagation conditions as well as boosts energy efficiency and spectral efficiency \cite{Larsson2014,Yang2015a}. It is widely recognized that the use of frequency division duplex (FDD) in massive MIMO incurs a significant channel estimation burden. To relieve this burden, the time-division duplex (TDD) mode is preferred to be used together with massive MIMO over the FDD mode \cite{Rusek2013,Lu2014}. In the TDD mode, the uplink and the downlink share the same frequency band such that the channel estimated through the uplink can be utilized in the downlink transmission \cite{Zhang2015}.

\subsection{Background and Motivation}

Pilot contamination has been identified as one of the key challenges to unlock the full potential of massive MIMO \cite{Zhu2015A,Khansefid2015,Khansefid2015b}. Principally, pilot contamination occurs when non-orthogonal pilot sequences are assigned to users. In typical massive MIMO networks, a very large number of users are served but only a limited number of orthogonal pilot sequences are available. The most serious consequence of pilot contamination is that it corrupts the channel estimates obtained at the BS during the uplink training phase, which results in an increased interference in the network. As such, pilot contamination is a commonly encountered problem and a major performance limiting factor in massive MIMO networks.

Some significant efforts have been devoted to address the pilot contamination problem in massive MIMO networks. These efforts are referred to as conventional methods and classified into four broad categories \cite{Lu2014}: 1) the protocol-based method, which restricts the simultaneous transmission from the users having the same pilot sequence or wisely assigns pilot sequences among users to alleviate pilot contamination (e.g., \cite{Fernandes2013,Zhu2015B,Ahmadi2015}); 2) the precoding-based method, which uses precoders to reduce the interference caused by pilot contamination (e.g., \cite{Jose2011}); 3) the angle-of-arrival (AoA)-based method, which mitigates the interference from the users having the same pilot sequence and mutually non-overlapping AoA (e.g., \cite{Yin2013}); and 4) the blind method, which partitions the signal space into desired signal subspace and interference signal subspace and then develops algorithms to reduce the interference from the latter (e.g., \cite{Muller2014}). It is worth mentioning that most conventional methods, e.g., \cite{Fernandes2013,Jose2011,Yin2013}, assumed orthogonal pilot sequences to perform pilot contamination analysis.

Recently, increasing attention has been paid to an alternative method, i.e., pilot sequence design, which aims at directly mitigating the detrimental impact of pilot contamination on the performance of massive MIMO networks. For example, when the number of users in the massive MIMO network exceeds the length of pilot sequence, \cite{Wang2015} designed pilot sequences by solving a Grassmannian manifold line packing problem. Different from \cite{Wang2015}, \cite{So2015} designed pilot sequences by solving an optimization problem aiming at maximizing the received signal-to-noise ratio. We note that neither \cite{Wang2015} nor \cite{So2015} considered the signal-to-interference-plus-noise ratio (SINR) requirements of individual users in designing pilot sequences. Hence, the results in \cite{Wang2015,So2015} cannot be used to evaluate the user capacity, which is a pivotal performance indicator of massive MIMO networks. User capacity is defined as the number of users that are admissible in a massive MIMO network such that the pre-defined SINR requirements of all the users in the network are satisfied. Against this background, \cite{Jung2013} evaluated the user capacity of a single-cell multiuser massive MIMO network. Taking the effective bandwidth into account, \cite{Zhong2007} explored the feasibility conditions of the same network. We note that the limitation of \cite{Jung2013,Zhong2007} lies in the assumption of perfect channel knowledge at the BS. This assumption does not generally hold in practice. To overcome this limitation, \cite{Shen2015} proposed a user capacity-achieving pilot sequence design together with power allocation for downlink transmission of a single-cell multiuser massive MIMO network. While \cite{Shen2015} laid a solid foundation to design pilot sequences in the single-cell network, the design of user capacity-achieving pilot sequences in the multi-cell multiuser massive MIMO network has not been explored in the literature. Different from \cite{Shen2015}, in this work we consider a more general multi-cell network, which encompasses the single-cell network as a special case. Moreover, our work provides a specific method for downlink power allocation and proves the feasibility of designing user capacity-achieving pilot sequences in multi-cell network even in the presence of inter-cell interference. Furthermore, we derive an expression for the limit on the maximum permitted SINR in a cell, which is not straightforward to obtain from \cite{Shen2015}.

\subsection{Contributions and Novelty}

In this paper, we mitigate pilot contamination in a multi-cell multiuser massive MIMO network by designing user capacity-achieving pilot sequences with low correlation coefficients. Crucially, our proposed design generates pilot sequences for each cell independently. The generated pilot sequences and the corresponding power allocation scheme satisfy the SINR requirements of all users in the network, regardless of the severity of pilot contamination. Our design is based on the rules of the generalized Welch-bound-equality (GWBE) sequence design\footnote{The GWBE sequence has been shown to achieve the user capacity in overloaded code-division-multiple-access (CDMA) systems \cite{Viswanath1999a,Viswanath1999}. We note that most studies on the signature sequence design in CDMA systems considered a single cell, e.g., \cite{Ulukus2001,Tropp2004}. Even when the multi-cell CDMA system was considered, e.g., \cite{Kishore2006}, the focus was to investigate hard hand-offs, rather than signature sequence design. This indicates the novelty of designing GWBE pilot sequences in multi-cell multiuser massive MIMO networks.}. The contributions and novelty of this work are summarized as follows:
\begin{itemize}
\item
We derive a new closed-form expression for the user capacity of the multi-cell multiuser massive MIMO network. This expression explicitly reveals that the user capacity is limited by the effective bandwidth\footnote{The effective bandwidth is a concept adopted in CDMA systems (e.g., \cite{Viswanath1999a,Zhong2007}). It represents a fraction of available degrees of freedom (i.e., the pilot sequence length) required by a user to achieve its target SINR. A user having a high SINR requirement requires a large fraction of available degrees of freedom and accordingly a large effective bandwidth. We clarify that the effective bandwidth is different from the channel bandwidth.}, the length of pilot sequence, the number of cells, and the number of users in each cell. Based on this expression, we derive a simple yet valuable result to determine the user capacity region of the network, under which the SINR requirements of all users are always satisfied.
\item
We propose a new algorithm to generate user capacity-achieving pilot sequences based on the derived user capacity region. The pilot sequences are independently generated for each cell. Thus, the proposed algorithm requires no cooperation among BSs. We also devise the power allocation scheme to control the downlink transmit power at each BS. This scheme only requires the BSs to exchange the correlation coefficient between pilot sequences with each other.
\item
We determine the minimum number of antennas required at BSs to achieve certain SINR requirements of all the users. This result is of practical significance since it avoids the uneconomic hardware costs caused by using unnecessary antennas to deliver the required quality of service.
\item
We analytically compare the performance of our design with that of two existing pilot sequence designs, namely, the Welch-bound-equality (WBE) design and the finite orthogonal set (FOS) design. Based on the analysis, we demonstrate that our design achieves a larger user capacity region than the existing designs.
\end{itemize}

Beyond the aforementioned contributions, we undertake a series of numerical evaluations to offer practically important insights into our design. The numerical results confirms that our GWBE design fulfills the SINR requirements of all the users throughout the network, while the existing designs are unable to fulfill the same SINR requirements.

\emph{Notations}: Vectors and matrices are denoted by lower-case and upper-case boldface symbols, respectively. $(\cdot)^{T}$ denotes the transpose, $(\cdot)^{H}$ denotes the Hermitian transpose, $\otimes$ denotes the Kronecker product, $\textrm{card}(\cdot)$ denotes the cardinality, $\mathbb{E}[\cdot]$ denotes the mathematical expectation, $\mathbf{I}_{M}$ denotes an $M\times M$ identity matrix, $\|\cdot\|$ denotes the ${l_{2}}$ norm, $\textrm{tr}(\cdot)$ denotes the trace operation, and $\text{var}(\cdot)$ denotes the variance operation.

\section{Multi-Cell Multiuser Massive MIMO Networks with Pilot Contamination}\label{System_Model}

\begin{figure}[!t]
\centering
\includegraphics[width=19pc]{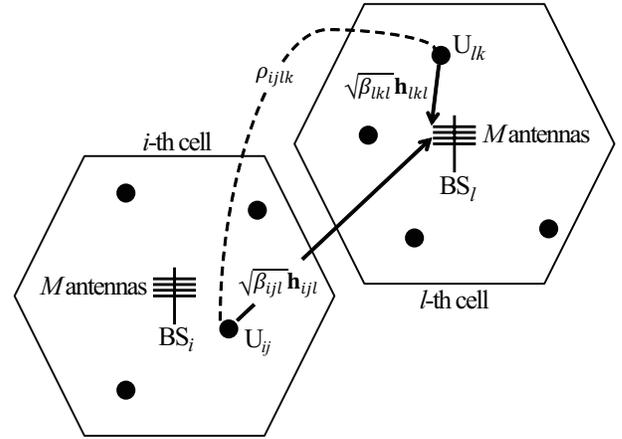}
\caption{Illustration of a TDD multi-cell multiuser massive MIMO network with $L = 2$, $K = 4$, and $K_{\textrm{tot}} = 8$. The uplink channel estimate is corrupted by pilot contamination when non-orthogonal pilot sequences are used in the network, i.e., $\rho_{ijlk}\neq{}0$.}\label{system_diag}
\end{figure}

We consider a TDD multi-cell multiuser massive MIMO network, an example of which is depicted in Fig.~\ref{system_diag}. The network consists of $L$ cells. In each cell, a BS equipped with $M$ antennas communicates with $K$ users, equipped with a single antenna each. We denote $K_{\textrm{tot}}$ as the total number of users in the network, where $K_{\textrm{tot}}=KL$. We also denote $\textrm{BS}_{l}$ as the BS in the $l$-th cell, where $l\in\left\{1,\dotsc,L\right\}$. We further denote $\textrm{U}_{ij}$ as the $j$-th user in the $i$-th cell, where $i\in\left\{1,\dotsc,L\right\}$ and $j\in\left\{1,\dotsc,K\right\}$.

We consider a block fading channel model, where the channel remains constant during the coherence time interval, $T$, and changes independently every interval. We denote $h_{ijlm}$ as the small-scale fading coefficient from $\textrm{U}_{ij}$ to the $m$-th antenna at $\textrm{BS}_{l}$, where $m\in\left\{1,\dotsc,M\right\}$. We assume that the small-scale fading coefficients in the network are subject to independent and identically distributed (i.i.d) Rayleigh fading. This assumption is reasonable in practice since the i.i.d Rayleigh fading model has recently been shown to agree well with the experimental data in massive MIMO \cite{Gao2015}. Based on this assumption, we find that $h_{ijlm}$ follows a complex Gaussian distribution with zero mean and unit variance, i.e., $h_{ijlm}\sim\mathcal{CN}(0,1)$. We next denote $\beta_{ijl}$ as the large-scale propagation factor from $\textrm{U}_{ij}$ to $\textrm{BS}_{l}$, which captures the effects of path loss and shadowing \cite{Jose2011}. Based on $\beta_{ijl}$ and ${h}_{ijlm}$, the overall propagation factor from $\textrm{U}_{ij}$ to \textit{m}-th antenna at $\textrm{BS}_{l}$ is given by $\sqrt{\beta_{ijl}}{h}_{ijlm}$. We highlight that the multi-cell multiuser massive MIMO network considered in this paper is a generalized model that describes a wide range of cell layouts.

Throughout this paper, we assume that the network operates in the TDD mode, where the uplink and the downlink channels are reciprocal \cite{Yan2015,Kan2014}. We also assume that the uplink training and downlink transmission time is less than or equal to the channel coherence interval.

\subsection{Uplink Training}

We first focus on the uplink training phase, during which the users in each cell send pilot sequences to enable channel estimation at the same-cell BS. At the beginning of each coherence interval, the $K$ users in the $l$-th cell send their pilot sequences of length $\tau$ to $\textrm{BS}_{l}$ in pre-assigned time slots. Assuming perfect synchronization, the pilot sequence vector received at $\textrm{BS}_{l}$ during the uplink training phase, denoted by a $\tau M \times 1$ vector, $\mathbf{y}_{l}$, is given by
\begin{align}\label{rec_pilot}
\mathbf{y}_{l}=\sum_{i=1}^{L}\sum_{j=1}^{K}\sqrt{p_{ij}\beta_{ijl}}{\mathbf{S}_{ij}}\mathbf{h}_{ijl}+\mathbf{z}_{l},
\end{align}
where $p_{ij}$ is the pilot power at $\textrm{U}_{ij}$, $\mathbf{S}_{ij} = {\mathbf{s}_{ij}}\otimes \mathbf{I}_{M}$ is a $\tau M \times M$ matrix, $\mathbf{s}_{ij}$ is a $\tau\times1$ pilot sequence assigned to $\textrm{U}_{ij}$, $\mathbf{h}_{ijl}=[{h}_{ijl1},{h}_{ijl2},\dotsc,{h}_{ijlM}]^{T}$ is the  $M\times1$ uplink  channel vector from $\textrm{U}_{ij}$ to $\textrm{BS}_{l}$, and $\mathbf{z}_{l}\sim\mathcal{CN}(0,\sigma_{z_{l}}^2)$ is the $\tau M \times 1$ additive white Gaussian noise (AWGN) at $\textrm{BS}_{l}$.

We assume that the pilot sequences are real and have unit energy. We denote $\rho_{ijlk}=\mathbf{s}_{lk}^{T}\mathbf{s}_{ij}$ as the correlation coefficient between different pilot sequences, where $k\in\left\{1,\dotsc,K\right\}$. The value of $\rho_{ijlk}$ varies from $-1$ to $+1$. Here, $+1$ indicates a perfect positive correlation coefficient, $-1$ indicates a perfect negative correlation coefficient, and $0$ indicates orthogonal pilot sequences with no correlation. As such, the pilot sequences matrix $\mathbf{S}_{ij}$ has a useful property, i.e., $\mathbf{S}_{ij}^{T}\mathbf{S}_{ij}=\mathbf{I}_{M}$. Of course, the ideal value for $\rho_{ijlk}$ is zero. However, this ideal value is not achievable in practical multiuser massive MIMO networks, which gives rise to pilot contamination.

The uplink channel vector $\mathbf{h}_{lkl}$ between $\textrm{U}_{lk}$ and $\textrm{BS}_{l}$ is estimated at $\textrm{BS}_{l}$ based on $\mathbf{y}_{l}$. We assume that the low-complexity least square (LS) channel estimation is performed to obtain an estimate $\mathbf{\hat{h}}_{lkl}$ for the uplink channel vector $\mathbf{h}_{lkl}$. It has been demonstrated that mean square error of an LS channel estimator remains nearly constant as $M$ increases \cite{Khansefid2015}, which makes it attractive for massive MIMO networks. As per the rules of the LS channel estimation, $\mathbf{\hat{h}}_{lkl}$ is given by
\begin{align}\label{channel_estimate}
\mathbf{\hat{h}}_{lkl} = \mathbf{S}_{lk}^{T}\mathbf{y}_{l}=  {\mathbf{S}_{lk}^T} \left( \sum_{i=1}^{L}  \sum_{j=1}^{K}  \sqrt{p_{ij}\beta_{ijl}} {\mathbf{S}_{ij}} \mathbf{h}_{ijl} + \mathbf{z}_{l} \right).
\end{align}
Moreover, we consider that the uplink power control is enabled such that $p_{lj}\beta_{ljl}=1$ \cite{Shen2015}. Accordingly, we re-express \eqref{channel_estimate} as
\begin{align}\label{channel_estimate_2}
\mathbf{\hat{h}}_{lkl} =  \mathbf{h}_{lkl}+ \sum_{i,j \neq l,k}  \rho_{ijlk} \sqrt{p_{ij}\beta_{ijl}}\mathbf{h}_{ijl} + {\mathbf{S}_{lk}^T} \mathbf{z}_{l},
\end{align}
where ${\sum}_{i,j\neq{}l,k}=\sum_{i=1}^{L}\sum_{j=1}^{K}$ and $(i,j) \neq (l,k)$.%

The effect of pilot contamination on the channel estimate can be easily seen from \eqref{channel_estimate_2}. When $\rho_{ijlk} \neq 0$, i.e., the correlation coefficient between the pilot sequences assigned to different users is non-zero, the channel estimate $\mathbf{\hat{h}}_{lkl}$ is contaminated by undesired channels $\mathbf{h}_{ijl}$.

\subsection{Downlink Transmission}

We now focus on the downlink transmission phase, during which the BS in each cell sends data symbols to the \textit{K} same-cell users. We denote $x_{lk}$ as the uncorrelated data symbols with zero mean. The transmit power of the data symbol $x_{lk}$ at $\textrm{BS}_{l}$ is given by $\mathbb{E} \left[x_{lk}^H {x_{lk}} \right] = P_{lk}$.

We consider that the data symbols for the downlink transmission are precoded by a linear precoding vector $\mathbf{t}_{lk}$. It has been acknowledged that the use of low-complexity linear precoding is preferred over high-complexity non-linear precoding in massive MIMO networks, which is due to the fact that linear precoding provides a near optimal performance \cite{Lu2014}. As such, we focus on a widely-adopted linear precoding scheme, maximum-ratio-transmission (MRT) \cite{Rusek2013}, in this work. The MRT precoding vector for $\textrm{U}_{lk}$ is given by $\mathbf{t}_{lk}=\frac{\mathbf{\hat{h}}_{lkl}}{\|\mathbf{\hat{h}}_{lkl}\|}$. We note that $\mathbf{t}_{lk}$ can be further simplified by utilizing the channel hardening property of massive MIMO networks \cite{Narasimhan2014}. This property implies that the channels between the BS and the $K$ users become increasingly orthogonal to each other when $M\rightarrow\infty$, which is given by \cite{Shen2015}
\begin{align}\label{ortho_mm}
\frac{1}{M}\mathbf{h}_{ijl}^{H}\mathbf{h}_{lkl}=
\begin{cases}
1, & \forall~(i,j)=(l,k)\\
0, & \text{otherwise.}
\end{cases}
\end{align}
Using \eqref{channel_estimate_2} and \eqref{ortho_mm}, we rewrite $\mathbf{t}_{lk}$ as
\begin{align}\label{precoding_vec}
\mathbf{t}_{lk}=\frac{\mathbf{\hat{h}}_{lkl}}{\sqrt{M\alpha_{lk}}},
\end{align}
where $\alpha_{lk}=\sum_{i=1}^{L} \sum_{j=1}^{K}\rho_{ijlk}^{2}\Xi_{ijl}^2+\sigma_{z}^{2}$ and  $\Xi_{ijl}\triangleq \sqrt{p_{ij}\beta_{ijl}}$.

By applying the MRT precoding at $\textrm{BS}_{l}$, the received signal at $\textrm{U}_{lk}$ through the downlink transmission phase is given by
\begin{align}\label{rec_initial}
\hat{r}_{lk} =  \sum_{m=1}^{L} \sum_{n=1}^{K}  \sqrt{\beta_{lkm}} {\mathbf{h}_{lkm}^H}\left( \mathbf{t}_{mn}x_{mn} \right) + w_{lk}
\end{align}
where $w_{lk}$ is the AWGN at $\textrm{U}_{lk}$. We assume that only the statistical information, $\mathbb{E}\left[\mathbf{g}_{lk}\right]$, is available at $\textrm{U}_{lk}$, where $\mathbf{g}_{lk}=\mathbf{h}_{lkl}^H \mathbf{t}_{lk}$. We clarify that this assumption is reasonable, since $\mathbf{g}_{lk}\approx\mathbb{E}\left[\mathbf{g}_{lk}\right]$ can be found based on of the channel hardening property in massive MIMO networks. %Moreover, acquiring the CSI at the user will require additional downlink pilots.
Relying on this assumption, we rewrite \eqref{rec_initial} as
\begin{align}\label{received_sig}
\hat{r}_{lk}=\sqrt{\beta_{lkl}} \mathbb{E} \left[{\mathbf{g}_{lk}}\right]x_{lk}+a_{lk},
\end{align}
where
\begin{align}\label{noise}
a_{lk}&=\sqrt{\beta_{lkl}}\left({\mathbf{g}_{lk}}-\mathbb{E} \left[{\mathbf{g}_{lk}}\right]\right)x_{lk} \notag \\ &+ \sum_{m,n \neq l,k} \sqrt{\beta_{lkm}}\mathbf{h}_{lkm}^H \left(\mathbf{t}_{mn}x_{mn}\right)  + w_{lk}.
\end{align}
We clarify that ${a}_{lk}$ in \eqref{received_sig} is treated as the effective noise and is uncorrelated with $\sqrt{\beta_{lkl}} \mathbb{E} \left[{\mathbf{g}_{lk}}\right] x_{lk}$.

\subsection{SINR at Users}

We now determine the achievable SINR from the $M$ antennas at $\textrm{BS}_{l}$ to $\textrm{U}_{lk}$, denoted by $\theta_{lk,M}$. The achievable SINR allows us to evaluate the ergodic achievable rate for $\textrm{U}_{lk}$, given by $R_{lk}= \log_{2}\left(1+\theta_{lk,M}\right)$ \cite{Jose2011,Shen2015}.

Based on \eqref{received_sig}, $\theta_{lk,M}$ is expressed as
\begin{align}\label{long_exp}
\theta_{lk,M}=\frac{\left(\mathbb{E} \left[{\mathbf{g}_{lk}} \right]\right)^2\beta_{lkl}P_{lk}}{\text{var}\left[{\mathbf{g}_{lk}} \right]\beta_{lkl}P_{lk} + \bar{\theta}_{lk,M} + \sigma_{w}^2},
\end{align}
where
\begin{align}\label{long_exp}
\bar{\theta}_{lk,M}\sum_{m,n \neq l,k} \mathbb{E}\left[\left|{\mathbf{g}_{lk}^{mn}} \right|^2\right]\beta_{lkm}P_{mn},
\end{align}
and $\mathbf{g}_{lk}^{mn}=\mathbf{h}_{lkm}^{H}\mathbf{t}_{mn}$.

We clarify that the achievable SINR given by \eqref{long_exp} is a generalized expression since it is valid for any precoder. We next specify the achievable SINR for the MRT precoding with LS channel estimation.
\begin{lemma}\label{lemma_red}
If the MRT precoding is used with the LS channel estimation, the SINR is derived as
\begin{align}\label{SINR}
\theta_{lk,M}=\frac{\beta_{lkl}P_{lk}}{\alpha_{lk}\left(\sum_{m,n \neq l,k}  \frac{\rho_{lkmn}^2\Xi_{lkm}^{2}\beta_{lkm}P_{mn}}{\alpha_{mn}}\right)+ \frac{\alpha_{lk}}{M}\left( \bar{P}_{lk} \right)},
\end{align}
where $\bar{P}_{lk} = \sum_{m,n}\beta_{lkm}P_{mn} + \sigma_{w}^2$.
\begin{IEEEproof}
Please refer to Appendix \ref{SINR_proof}.
\end{IEEEproof}
\end{lemma}

We next draw some valuable insights from \eqref{SINR}. First, \eqref{SINR} is easy to compute since it is independent of channel realizations. Second, \eqref{SINR} captures the effect of correlation coefficient between pilot sequences on the achievable SINR. Finally, \eqref{SINR} shows that the achievable SINR $\theta_{lk,M}$ increases with $M$.

Next, we provide an asymptotic expression for the achievable SINR in \eqref{SINR} when the number of antennas at the BS grows very large, i.e., $M\rightarrow\infty$. We note that $M\rightarrow\infty$ is a valid assumption in massive MIMO networks. Under this assumption, we derive the asymptotic achievable SINR, $\theta_{lk,\infty}$, as
\begin{align}\label{SINR_inf}
\theta_{lk,\infty}&=
\frac{\beta_{lkl}P_{lk}}{\alpha_{lk}\left(\sum_{m,n}  \frac{\rho_{lkmn}^2\Xi_{lkm}^{2}\beta_{lkm}P_{mn}}{\alpha_{mn}}\right)- \beta_{lkl}P_{lk}}.%=\frac{P_{lk}}{\alpha_{lk}\textrm{tr} \left(\mathbf{s}_{lk}^{T} \mathbf{S}\mathbf{P}\mathbf{A}_{lk} \mathbf{S}^{T}\mathbf{s}_{lk}\right)-P_{lk}},
\end{align}

It is seen from \eqref{SINR_inf} that pilot contamination always limits the achievable SINR, even when $M\rightarrow \infty$. It also highlights the severity of the pilot contamination problem in massive MIMO networks.

\section{User Capacity Analysis and Pilot Sequence Design}\label{sec:design}

In this section, we first derive the user capacity of the multi-cell multiuser massive MIMO network. We then detail the sufficient condition for achieving the maximum user capacity by designing a set of user capacity-achieving pilot sequences for the users in each cell. Furthermore, we
determine the minimum number of antennas at BSs to achieve the given SINR requirements of all the users in the network.

\subsection{User Capacity and Its Condition}\label{sec:user_capacity}

Throughout this paper, we define the user capacity as the maximum number of users that can be served simultaneously via downlink transmission such that $\textrm{U}_{ij}$ meets the predefined SINR requirement, denoted by $\gamma_{ij}$. We focus on a practical scenario of the multi-cell multiuser massive MIMO network where $K_{\textrm{tot}}>K>\tau$. In this scenario, the pilot contamination is caused by both inter-cell pilot sequences and intra-cell pilot sequences\footnote{We note that the intra-cell pilot contamination can be avoided in the pilot sequence design if $K_{\textrm{tot}}>\tau>K$. However, the condition $K<\tau$ imposes a limitation on the value of $K$. As such, in this work we consider the scenario with $K_{\textrm{tot}}>K>\tau$ such that both inter-cell pilot contamination and intra-cell pilot contamination occur.}. As such, we need to address both inter-cell pilot contamination and intra-cell pilot contamination in the user capacity analysis and pilot sequence design. We derive the user capacity, i.e., $K_{\textrm{tot}}$ of a multi-cell multiuser massive MIMO network in the following proposition:
\begin{proposition}\label{prop_K_tot}
In an $L$-cell multiuser massive MIMO network, $K_{\textrm{tot}}$ users can be served simultaneously through the downlink transmission such that the SINR requirements of all the users in the network are satisfied if
\begin{align}\label{K_tot}
K_{\textrm{tot}}\leq\sqrt{\tau\sum\limits_{i=1}^{L}\sum\limits_{j=1}^{K}
\frac{1+\gamma_{ij}}{\gamma_{ij}}}.
\end{align}
\begin{IEEEproof}
Please refer to Appendix \ref{prop1_proof}.
\end{IEEEproof}
\end{proposition}

We highlight that \eqref{K_tot} determines the upper bound on the user capacity of the multi-cell multiuser massive MIMO network. As shown in \eqref{K_tot}, the upper bound on the user capacity depends on the effective bandwidth given by $\frac{\gamma_{ij}}{1+\gamma_{ij}}$, the length of pilot sequence $\tau$, the number of cells $L$, and the number of users in each cell $K$. More importantly, the bound on the user capacity implies the presence of a user capacity-achieving region under which the user capacity can be achieved. We next present the sufficient condition for achieving the bound on the user capacity in the following proposition:
\begin{proposition}\label{prop_2}
The bound on the user capacity of a multi-cell multiuser massive MIMO network is achieved when the sum of effective bandwidth of all the users in the network is less than or equal to the length of the pilot sequence. Mathematically, the condition is given by
\begin{align}\label{BW_all}
\sum_{i=1}^{L}\sum_{j=1}^{K}\frac{\gamma_{ij}}{1+\gamma_{ij}}\leq\tau.
\end{align}
\begin{IEEEproof}
Using the Cauchy-Schwarz inequality, we obtain
\begin{align}\label{proof_prop2}
\sum_{i=1}^{L}\sum_{j=1}^{K}\frac{1+\gamma_{ij}}{\gamma_{ij}}\geq\frac{K_{\textrm{tot}}^2}
{\sum_{i=1}^{L}\sum_{j=1}^{K}\frac{\gamma_{ij}}{1+\gamma_{ij}}}=\frac{K_{\textrm{tot}}^2}{\tau}.
\end{align}
Rearranging \eqref{proof_prop2} produces \eqref{K_tot}, thus completing the proof.
\end{IEEEproof}
\end{proposition}

We clarify that \eqref{BW_all} characterizes the upper bound on the user capacity region of the multi-cell multiuser massive MIMO network. Throughout this paper, we define the user capacity region as the region defined by a collection of admissible SINR tuples such that the upper bound given by \eqref{BW_all} is satisfied. In particular, the equality of \eqref{BW_all} gives the upper surface boundary of the user capacity region, i.e., $\sum_{i=1}^{L}\sum_{j=1}^{K}\frac{\gamma_{ij}}{1+\gamma_{ij}}=\tau$. We clarify that remaining inside this user capacity region is the sufficient condition for achieving the bound on the user capacity given by \eqref{K_tot}. Assuming that the user capacity region is equally shared among all cells in the network, the upper bound on the per-cell user capacity region is expressed as
\begin{align}\label{BW}
\sum_{j=1}^{K}\frac{\gamma_{ij}}{1+\gamma_{ij}} \leq \frac{\tau}{L}.
\end{align}
Using \eqref{BW}, the upper surface boundary of the per-cell user capacity region is obtained as $\sum_{j=1}^{K}\frac{\gamma_{ij}}{1+\gamma_{ij}} = \frac{\tau}{L}$. As such, the SINR requirements of all the users in a cell must satisfy the bound given by \eqref{BW} for a user capacity-achieving pilot sequence design. In our work, the motivation behind dividing the user capacity region equally among all the cells is to maintain fairness in the network\footnote{We note that the sum-rate is not necessarily the same for each cell in the network even when the user capacity region is equally divided among all the cells in the network. Specifically, the downlink power allocation is unique for each user, which affects the rate of the user. The impact of the proposed pilot sequence design on the rate symmetry between the cells is beyond the scope of our research and is left as future work.}. We highlight that the user capacity is fully characterized when the upper bound on the user capacity region of the network given by \eqref{BW_all} is satisfied, regardless of how the user capacity region is divided among all the cells.

We clarify that \eqref{K_tot}, \eqref{BW_all}, and \eqref{BW} represent the upper bound on the user capacity, the upper bound on the user capacity region, and the upper bound on the per-cell user capacity region, respectively, since they are derived for the large number of antenna regime, i.e., $M \to \infty$. Based on \eqref{BW_all} and \eqref{BW}, we now provide the following remarks on the pilot sequence design:
\begin{remark}
We find that the bound on the user capacity region given by \eqref{BW_all} is always satisfied when the bound on the per-cell user capacity region given by \eqref{BW} holds in each cell. This allows us to achieve the user capacity and design the user capacity-achieving pilot sequences by considering the per-cell SINR requirements independently and guaranteeing that \eqref{BW} is satisfied. As such, we focus on the per-cell user capacity region in designing the desired pilot sequences.
\end{remark}

\begin{remark}
The inequality \eqref{K_tot} can help network designers to identify the appropriate network parameters, i.e., $\tau$,$~K$,$~L$, and $\gamma_{ij}$ to ensure that the network is able to serve the required number of users, i.e, $K_{\textrm{tot}}$. Once the parameters $\tau$,$~K$,$~L$ are given, the inequality \eqref{BW_all} provides additional condition on $\gamma_{ij}$ to guarantee that $K_{\textrm{tot}}$ user are admissible in the network such that their SINR requirement are met.
\end{remark}

\subsection{Pilot Sequence Design}\label{sec:pilot_design}

In this subsection, we develop a new algorithm to design the user capacity-achieving pilot sequences. Here, the user capacity-achieving pilot sequences are defined as the pilot sequences satisfying the SINR requirements of all the users in the network and achieving the bound on the per-cell user capacity region given by \eqref{BW} (or equivalently, achieving the bound on the user capacity region given by \eqref{BW_all}). We highlight that our preliminary pilot sequence design scheme was examined in one of our previous works \cite{Akbar2016}.

We first present three preliminaries that aid in the pilot sequence design. To this end, we define two $1\times{}K$ vectors $\mathbf{z}$ and $\mathbf{x}$ as $\mathbf{z}=\left[\frac{\gamma_{l1}}{1+\gamma_{l1}}, \frac{\gamma_{l2}}{1+\gamma_{l2}}, \dotsc, \frac{\gamma_{lK}}{1+\gamma_{lK}}\right]$ and $\mathbf{x}=\left[x_{1},x_{2},\dotsc,x_{\tau},0,\dotsc,0\right]$, respectively, where $\gamma_{l1}\geq\gamma_{l2}\geq\dotsc \geq\gamma_{lK}$ and $x_{1}\geq x_{2}\geq \dotsc x_{\tau}$. We emphasize that the SINR requirements $\gamma_{lk}$ need to be carefully chosen such that \eqref{BW} is satisfied. The three preliminaries based on $\mathbf{x}$ and $\mathbf{z}$ are given as follows:
\newtheorem{preliminary}{Preliminary}
\begin{preliminary}\label{prem1}
Given vectors $\mathbf{z}$ and $\mathbf{x}$, $\mathbf{x}$ majorizes $\mathbf{z}$, i.e., $\mathbf{x}\succ\mathbf{z}$, if $\sum_{n=1}^{m}x_{n}\geq\sum_{n=1}^{m}z_{n}$, where $m\in\left\{1,\cdots,K-1\right\}$ and $\sum_{n=1}^{m}x_{n}=\sum_{n=1}^{m}z_{n}$, where $m=K$.
\end{preliminary}
\begin{preliminary}\label{prem2}
Given a vector $\mathbf{z}$, a vector $\mathbf{x}$ can be found for the value of $m=\tau$ such that  $\mathbf{x}\succ\mathbf{z}$, if $\mathbf{x}$ is given by $x_{i}=\sum_{n=1}^{K}\frac{z_{n}}{\tau}$, where $i\in\left\{1,\cdots,\tau\right\}$ and $x_{j}=0$, where $j\in\left\{\tau+1,\cdots,K\right\}$.
\end{preliminary}
\begin{preliminary}\label{prem3}
If $\mathbf{x}\succ\mathbf{z}$, $\mathbf{z}$ is obtainable by applying at most $K-1$ T-transform operations \cite{Hardy1954} on $\mathbf{x}$, i.e., $\mathbf{z}=\mathbf{T}_{K-1}\mathbf{T}_{K-2}\cdots\mathbf{T}_{1}\mathbf{x}$, and there exists a matrix $\mathbf{W}=\mathbf{W}_{1}\mathbf{W}_{2}\cdots\mathbf{W}_{K-1}$, where $\mathbf{W}_{i}$ is a unitary matrix generated from $\mathbf{T}_{i}$ at each step of the T-transform \cite{Hardy1954,Viswanath1999}.
\end{preliminary}

\emph{Preliminary \ref{prem1}} is based on the majorization theory \cite{Marshall2011}. Recently, the majorization theory has been used to find fundamental limits on the throughput of a multi-user MIMO system \cite{Yuan2016}. Different from \cite{Yuan2016}, where the majorization theory is used to derive the upper bound and lower bound on the system throughput, we utilize the majorization theory to design user capacity-achieving pilot sequences for a multi-cell multi-user massive MIMO network.

\begin{algorithm}
\caption{User capacity-achieving pilot sequence design} \label{algo1}
\begin{algorithmic}[1]
\Statex
\Statex
\Function{Pilot Design}{$\mathbf{\Gamma},\tau$}\Comment{$\mathbf{\Gamma} = \left[\pmb{\gamma}_{1},\cdots,\pmb{\gamma}_{L}\right]$}
\For{$l\gets 1, L$}
\State $\pmb{\gamma} \gets  \pmb{\gamma}_{l}$
\Comment{\parbox[t]{.6\linewidth}{$\pmb{\gamma}$ is set to the SINR requirements of the \textit{K} users in the \textit{l}th cell}}%
\State $\textrm{sum} \gets 0$
\State $\mathbf{x}_{l} = \mathbf{z}_{l}  \gets \mathbf{0}_{1\times K}$ \Comment{$\mathbf{x}_{l}$ is a $1 \times K$ zero vector}
\State $\hat{\pmb{\gamma}} \gets \Call{Gamma-Hat}{\pmb{\gamma}}$
\For{$k\gets 1, K$}
\State $\mathbf{z}_{l}(k)\gets \left(\frac{\hat{\gamma}_{k}}{1+\hat{\gamma}_{k}}\right)$ %\Comment{The element, i.e., $\mathbf{z}_{l}(k)$ is set to the effective bandwidth of the $k$th user}
\State $\textrm{sum} \gets \textrm{sum} + \mathbf{z}_{l}(k)$
\EndFor
\State $B_{l} \gets \frac{\textrm{sum}}{\tau}$
\State $\mathbf{x}_{l}\left(1,\cdots,\tau\right) \gets B_{l}$ \Comment{\parbox[t]{.4\linewidth}{The first $\tau$ elements of $\mathbf{x}_{l}$ are set to $B_{l}$}}
\State $\mathbf{W}^{l} \gets \Call{T-transform}{\mathbf{z}_{l},\mathbf{x}_{l}}$
\State $\mathbf{V}_{l} \gets \mathbf{W}^{l}\left(\tau,:\right)$ \Comment{$\mathbf{V}_{l}$ retains first $\tau$ rows of $\mathbf{W}^{l}$}
\State $\mathbf{Z}_{l} \gets \text{diag}\{\mathbf{z}_{l}\}$ \Comment{$\mathbf{z}_{l}$ is a diagonal matrix}
\State $\mathbf{S}_{l} \gets \text{normc}\left(B_{l}^{\frac{1}{2}}\mathbf{V}_{l} \mathbf{Z}_{l}^{-\frac{1}{2}}\right)$
\EndFor
\State $\mathbf{S} \gets \left[\mathbf{S}_{1},\dotsc,\mathbf{S}_{L}\right]$ \Comment{Desired pilot sequence matrix}
\State \textbf{return} $\mathbf{S}$
\EndFunction
\Statex
\Statex
\Function{Gamma-Hat}{$\pmb{\gamma}$}
\For{$j\gets 1, K$}
\State \textrm{find} $\hat{\gamma}(j) \geq \gamma(j)$ \\
\hskip\algorithmicindent \hskip\algorithmicindent subject to $\sum_{q=1}^{K}\frac{\hat{\gamma}_{q}}{1+\hat{\gamma}_{q}}=\frac{\tau}{L}$\\
\hspace{2.6cm}and $\hat{\gamma}(j+1)\leq\hat{\gamma}(j)\leq\frac{1}{L-1}$.
\EndFor
\State \textbf{return} $\hat{\pmb{\gamma}}$
\EndFunction
\Statex
\Statex
\Function{T-transform}{$\mathbf{z}_{l},\mathbf{x}_{l}$}
\For{$i\gets 1, K-1$}
\State $\mathbf{T}_{i} = \mathbf{W}_{i} \gets \mathbf{I}_{K}$
\State $k_{\textrm{min}} \gets \min_{1\leq k\leq K} \left\{\mathbf{z}_{l}\left(k\right) < \mathbf{x}_{l}\left(k\right)\right\}$
\State $k_{\textrm{max}} \gets \max_{1\leq k\leq K} \left\{\mathbf{z}_{l}\left(k\right) > \mathbf{x}_{l}\left(k\right)\right\}$
\State $\xi \gets \frac{\min\left\{\mathbf{x}_{l}\left(k_{\textrm{min}}\right) - \mathbf{z}_{l}\left(k_{\textrm{min}}\right)~,~\mathbf{z}_{l}\left(k_{\textrm{max}}\right) - \mathbf{x}_{l}\left(k_{\textrm{max}}\right)\right\}}{\mathbf{x}_{l}\left(k_{\textrm{min}}\right)-\mathbf{x}_{l}\left(k_{\textrm{max}}\right)}$
\State $\mathbf{T}_{i}\left(k_{\textrm{min}},k_{\textrm{max}}\right) = \mathbf{T}_{i}\left(k_{\textrm{max}},k_{\textrm{min}}\right) \gets \xi$ \label{T_set1}
\State $\mathbf{T}_{i}\left(k_{\textrm{min}},k_{\textrm{min}}\right)=\mathbf{T}_{i}\left(k_{\textrm{max}},k_{\textrm{max}}\right) \gets 1-\xi$ \label{T_set2}
\For{$m\gets 1, K$} \label{U_set1}
\For{$n\gets 1, K$}
\If {$m\leq n$}
\State $\mathbf{W}_{i}\left(m,n\right)=\sqrt{\mathbf{T}_{i}\left(m,n\right)}$
\Else
\State $\mathbf{W}_{i}\left(m,n\right)=-\sqrt{\mathbf{T}_{i}\left(m,n\right)}$
\EndIf
\EndFor
\EndFor \label{U_set2}
\EndFor
\State $\mathbf{W}^{l} \gets \mathbf{W}_{1}\mathbf{W}_{2}\dotsc\mathbf{W}_{K-1}$
\State \textbf{return} $\mathbf{W}^{l}$
\EndFunction
\end{algorithmic}
\end{algorithm}

We next present \textbf{Algorithm~\ref{algo1}} for our proposed user capacity-achieving pilot design. Three functions are used in this algorithm, namely, \textsc{Pilot-Design}, \textsc{Gamma-Hat}, and \textsc{T-Transform}.  The functions utilize \emph{Preliminary \ref{prem1}}, \emph{Preliminary \ref{prem2}}, \emph{Preliminary \ref{prem3}}, and the results from Section III-A to design user capacity achieving pilot sequences.

\textbf{Algorithm~\ref{algo1}} describes the step-by-step process of designing the user capacity-achieving pilot sequences. The main function \textsc{Pilot-Design} takes two inputs: a $1\times K_{\textrm{tot}}$ vector, $\pmb{\Gamma}=\left[\pmb{\gamma}_{1},\pmb{\gamma}_{2},\dotsc,\pmb{\gamma}_{l},\dotsc,\pmb{\gamma}_{L}\right]$, and the length of pilot sequence, $\tau$. The $l$-th element in $\pmb{\Gamma}$ is given by a $1\times K$ vector, $\pmb{\gamma}_{l}=\left[\gamma_{l1},\gamma_{l2},\dotsc,\gamma_{lK}\right]$, which contains the SINR requirements of the $K$ users in the $l$-th cell, where $\gamma_{l1}\geq\gamma_{l2}\geq\dotsc\geq\gamma_{lK}$. At each time, the main function \textsc{Pilot-Design} considers the SINR requirements in one cell and returns $\mathbf{S}_{l}$ as the desired pilot sequence matrix for the users in the considered cell. Particularly, the $k$-th column in $\mathbf{S}_{l}$ is the pilot sequence for $\textrm{U}_{lk}$. We also clarify that the two functions \textsc{Gamma-Hat} and \textsc{T-Transform} are adopted in \textsc{Pilot-Design} to facilitate the pilot sequence design. Specifically, the function \textsc{Gamma-Hat} obtains $\hat{\gamma}\geq\gamma$ for each element of $\pmb{\gamma}$ to guarantee that the SINR requirements lie on the upper surface boundary of the per-cell user capacity region obtained from \eqref{BW}. The function \textsc{T-Transform} returns a $K\times K$ matrix $\mathbf{W}^{l}$, using \emph{Preliminary \ref{prem3}}, as a key enabler to obtain $\mathbf{S}_{l}$. As shown in \textbf{Algorithm~\ref{algo1}}, $\mathbf{V}^{l}$ is obtained from $\mathbf{W}^{l}$ and used to obtain $\mathbf{S}_{l}$ by normalizing the columns of $B_{l}^{\frac{1}{2}}\mathbf{V}_{l} \mathbf{Z}_{l}^{-\frac{1}{2}}$, represented by $\text{normc}\left(B_{l}^{\frac{1}{2}}\mathbf{V}_{l} \mathbf{Z}_{l}^{-\frac{1}{2}}\right)$.

We note that \textbf{Algorithm~\ref{algo1}} returns an effective pilot sequence matrix when \eqref{BW} is satisfied. We also note that the inequality given by \eqref{BW} may not hold if one user in a cell has a very high SINR requirement. This requires us to find the limit on the maximum permitted SINR requirement in a cell, which is given by $\gamma_{l}^{\textrm{MAX}}=\max_{1\leq k \leq K}\left(\gamma_{lk}\right)$. To this end, we first specify the condition for $\mathbf{x}_{l}\succ\mathbf{z}_{l}$ in the following Lemma.
\begin{lemma}\label{majorize}
$\mathbf{x}_{l}\succ\mathbf{z}_{l}$ if the maximum individual effective bandwidth, i.e., $z_{l}^{\textrm{MAX}}=\frac{\gamma_{l}^{\textrm{MAX}}}{1+\gamma_{l}^{\textrm{MAX}}}$, is less than or equal to $1/L$. %the reciprocal of the number of cells in the network.
\begin{IEEEproof}
Based on \eqref{BW}, the $i$-th element of $\mathbf{x}_{l}$ is given by
\begin{align} \label{proof_maj}
x_{i}=\frac{1}{\tau}\sum_{k=1}^{K}z_{k}=
\frac{1}{\tau}\sum_{k=1}^{K}\frac{\gamma_{lk}}{1+\gamma_{lk}}=
\frac{1}{\tau}\times\frac{\tau}{L}=\frac{1}{L}.
\end{align}
Since $\mathbf{x}_{l}\succ\mathbf{z}_{l}$, the largest element in $\mathbf{z}_{l}$ needs to be less than or equal to $x_{i}$, i.e., $z_{l}^{\textrm{MAX}}\leq{}x_{i}$. Using this inequality together with $x_{i}=\frac{1}{L}$, we have
\begin{align}\label{BW_ub}
\frac{\gamma_{l}^{\textrm{MAX}}}{1+\gamma_{l}^{\textrm{MAX}}} \leq \frac{1}{L}.
\end{align}
This completes the proof.
\end{IEEEproof}
\end{lemma}

\begin{corollary}
\label{corro123}
The limit on the maximum permitted SINR requirement in the $l$-th cell is obtained by simplifying \eqref{BW_ub} as
\begin{align}\label{idn_SINR}
\gamma_{l}^{\textrm{MAX}} \leq \frac{1}{L-1}
\end{align}
We clarify that \eqref{idn_SINR} rationalizes the condition of $\hat{\gamma}(j)\leq\frac{1}{L-1}$ used in the function \textsc{Gamma-Hat}.
\end{corollary}

We next demonstrate the effectiveness of the proposed pilot sequence design in the achievable SINR of $\textrm{U}_{lk}$ in the following proposition:
\begin{proposition}\label{prop_3}
When the pilot sequence is designed according to \textbf{Algorithm~\ref{algo1}} and the downlink power for each user is allocated proportional to the effective bandwith of the user, the SINR requirements of all the users in the network are satisfied.
\begin{IEEEproof}
We first make a reasonable assumption that the downlink transmit power for $\textrm{U}_{lk}$ at $\textrm{BS}_{l}$ needs to be chosen according to the SINR requirement of the particular user. Since the effective bandwidth, $\frac{\hat{\gamma}}{1+\hat{\gamma}}$, is adopted in the proposed pilot sequence design, the downlink transmit power for $\textrm{U}_{lk}$ at $\textrm{BS}_{l}$ is set to
\begin{align}\label{dp_alloc}
P_{lk}=\frac{\alpha_{lk}\hat{\gamma}_{lk}}{1+\hat{\gamma}_{lk}}.
\end{align}

The pilot sequences output by \textbf{Algorithm~\ref{algo1}} has the important property that $\mathbf{S}_{l}\mathbf{Z}_{l}\mathbf{S}_{l}^{T}=B_{l}\mathbf{I}_{\tau}$. Next, we use \eqref{BW_all}, \eqref{dp_alloc}, the matrix definitions given in Table \ref{matrix_definitions} in Appendix \ref{prop1_proof}, the pilot sequence property given by $\mathbf{S}_{l}\mathbf{Z}_{l}\mathbf{S}_{l}^{T}=B_{l}\mathbf{I}_{\tau}$, and the uplink power control assumption given by $p_{lk}\beta_{lkm} \leq 1$ to simplify \eqref{SINR_inf} as
\begin{align}\label{proof_ada_all}
\theta_{lk,\infty}&\geq\frac{P_{lk}}{\frac{\alpha_{lk}}{\tau}
\sum_{i,j}\frac{P_{ij}}{\alpha_{ij}}-P_{lk}}=\frac{\frac{\hat{\gamma}_{lk}}{1+\hat{\gamma}_{lk}}}{\frac{1}{\tau}\sum_{i,j} \frac{\hat{\gamma}_{ij}}{1+\hat{\gamma}_{ij}}
-\frac{\hat{\gamma}_{lk}}{1+\hat{\gamma}_{lk}}},\notag\\
&\geq \frac{\hat{\gamma}_{lk}}{1+\hat{\gamma}_{lk}-\hat{\gamma}_{lk}}=\hat{\gamma}_{lk}.
\end{align}
Given that $\hat{\gamma}_{lk}\geq\gamma_{lk}$, the SINR requirement of $\textrm{U}_{lk}$ is satisfied, which completes the proof.
\end{IEEEproof}
\end{proposition}

We refer to the designed user capacity-achieving pilot sequences as the GWBE pilot sequences. \emph{Proposition~\ref{prop_3}} shows that the the joint use of \textbf{Algorithm~\ref{algo1}} and the proposed downlink power allocation ensures that all the users in the network achieve their SINR requirements when \eqref{BW} is satisfied, although it designs pilot sequences for different cells separately. The motivation behind designing pilot sequences independently for each cell is to minimize overhead due to cooperation among BSs. To design pilot sequences with BS cooperation, our proposed design requires SINR feedback from all the users in the network. As such, the feedback overhead significantly increases with the number of users in the network. Additionally, we note that enabling cooperation among the BSs does not affect the user capacity of the network. This is because the user capacity is achieved as long as the bound on the user capacity given by \eqref{K_tot} holds. Our work shows that the bound on the user capacity can be achieved and the user capacity-achieving pilot sequences can be designed without cooperation among BSs, which implies little or no advantage of enabling cooperation among BSs in terms of achieving the user capacity.

We further highlight that \eqref{proof_ada_all} provides a key insight into choosing the value of $\hat{\gamma}$ in the function \textsc{Gamma-Hat}, i.e., $\hat{\gamma}_{lk}=\theta_{lk,\infty}$. Intuitively, choosing $\hat{\gamma}_{lk} = \theta_{lk,\infty}$ means that the SINR requirement for $\textrm{U}_{lk}$ can be satisfied with an infinite number of antennas at the BS. However, in practical scenarios, we have $\gamma_{lk} < \hat{\gamma}_{lk}$. As such, the SINR required for $\textrm{U}_{lk}$ is satisfied with a limited number of antennas at the BS with our pilot sequence design and downlink power allocation scheme.
\subsection{Minimum Number of Antennas for Required SINR}\label{sec:lim_anteana}

In this subsection, we determine the minimum number of antennas required at BSs to achieve certain SINR requirements, which helps to reduce the hardware costs incurred by using unnecessary antennas. We introduce $\mu$ as a performance satisfaction index, where $0<\mu<1$. Mathematically, $\mu$ is expressed as the ratio between the achievable SINR with finite $M$ and the achievable SINR with infinite $M$, i.e., $\mu=\theta_{lk,M}/\theta_{lk,\infty}$. The practical implication of introducing $\mu$ lies in its potential of enabling the network designers to find the desired $M$ to meet a proportion of the achievable SINR requirement with $M\rightarrow\infty$.

Using \eqref{SINR} and \eqref{SINR_inf}, the minimum $M$ required by $\textrm{U}_{lk}$ for our proposed GWBE pilot sequence design is given by
\begin{align}\label{antenna_gwbe}
M_{lk,\textrm{GWBE}}^{\textrm{MIN}} \geq \frac{\sum_{m,n} \beta_{lkm}P_{mn}+\sigma_{w}^{2}}{ \left(\frac{1-\mu}{\mu}\right){\sum}_{m,n\neq l,k}\frac{\rho_{lkmn}^{2}\Xi_{lkm}^2\beta_{lkm}P_{mn}}{\alpha_{mn}}}.
\end{align}
Based on \eqref{antenna_gwbe}, the minimum $M$ for the network is given by $M_{\textrm{GWBE}}^{\textrm{MIN}}=\max_{1\leq{}l\leq{}L,1\leq{}k\leq{}K}\{M_{lk,\textrm{GWBE}}^{\textrm{MIN}}\}$.
An important conclusion is reached from \eqref{antenna_gwbe} that the number of antennas at BSs needs to be at least $M_{\textrm{GWBE}}^{\textrm{MIN}}$, i.e., $M\geq{}M_{\textrm{GWBE}}^{\textrm{MIN}}$, to achieve the SINR requirements given by $\mu\theta_{lk,\infty}$.

\section{Comparison of Our Design with Other Designs}\label{sec:Comparison}

In this section, we compare the performance of our proposed GWBE pilot sequence design with the performance of the existing pilot sequence designs, namely, the WBE design \cite{Welch1974,Tropp2004} and the FOS design \cite{Jose2011,Marzetta2010}. This comparison allows us to demonstrate the advantage of our proposed design over the existing designs.

\subsection{Welch-Bound-Equality (WBE) Design}\label{sec:WBE_desc}
\begin{figure*}[!t]
\centering
\begin{minipage}{\textwidth}
\subfloat[$L=2$]{\includegraphics[width=14.1pc]{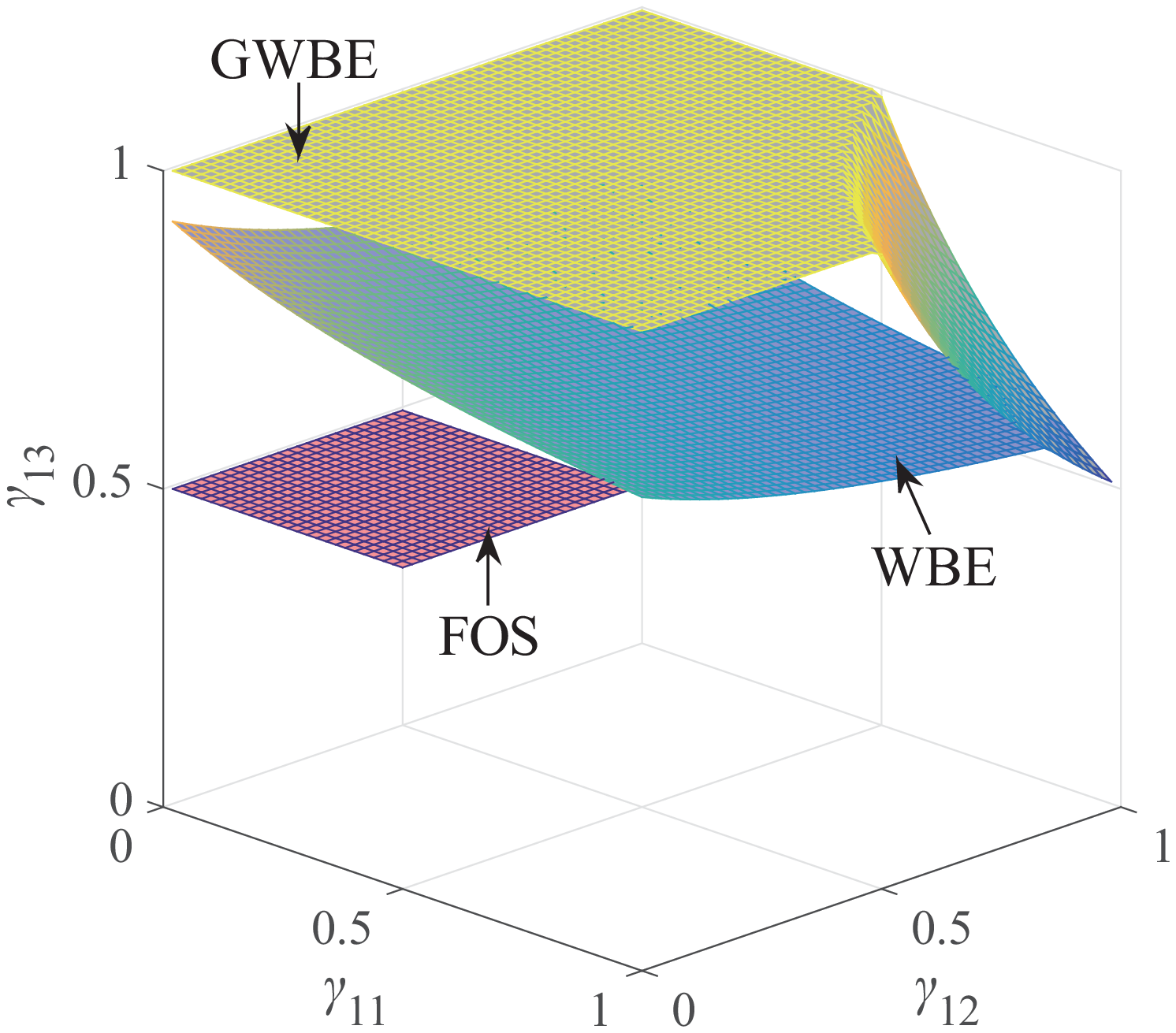}}
\subfloat[$L=3$]{\includegraphics[width=14.1pc]{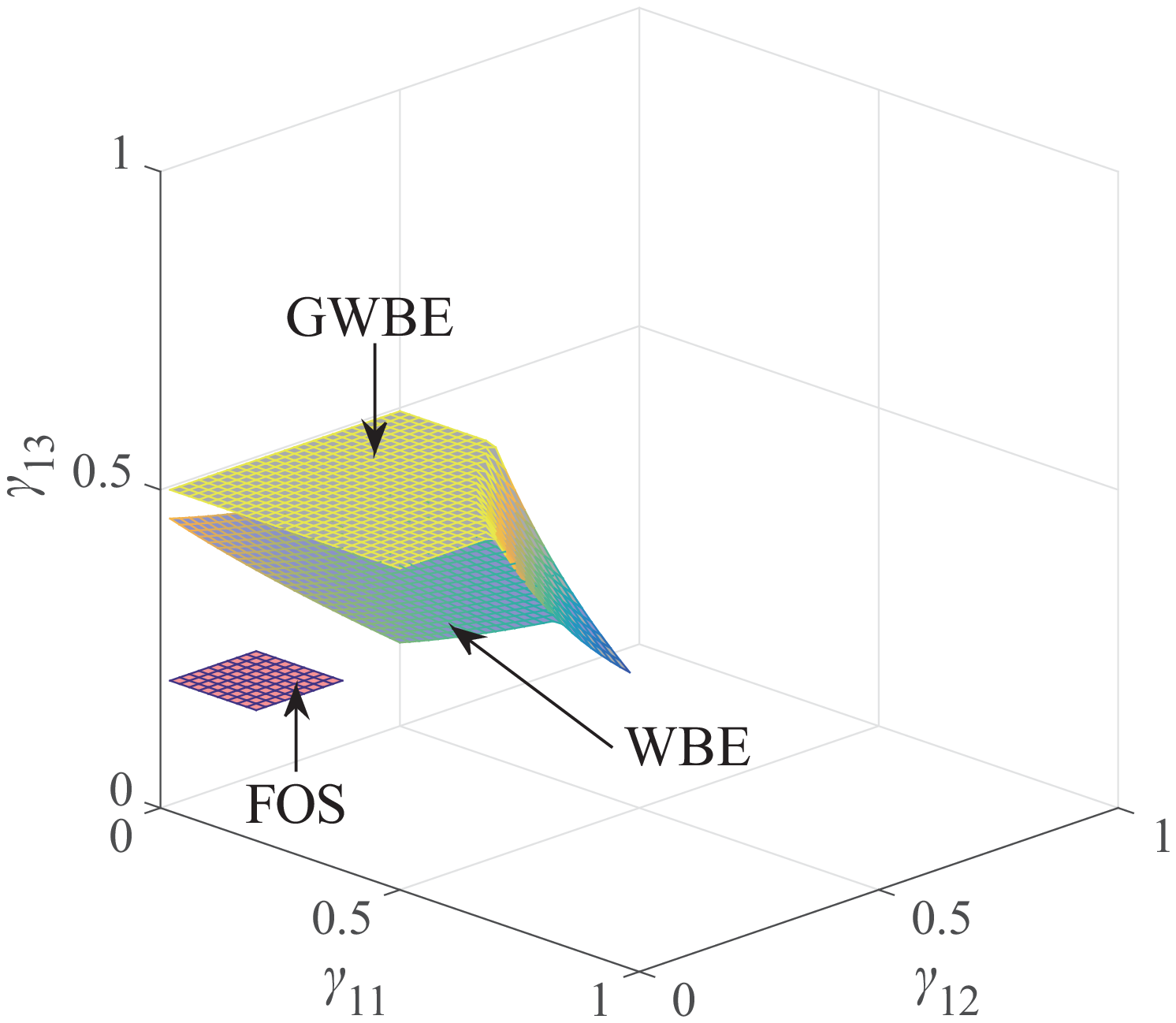}}
\subfloat[$L=4$]{\includegraphics[width=14.1pc]{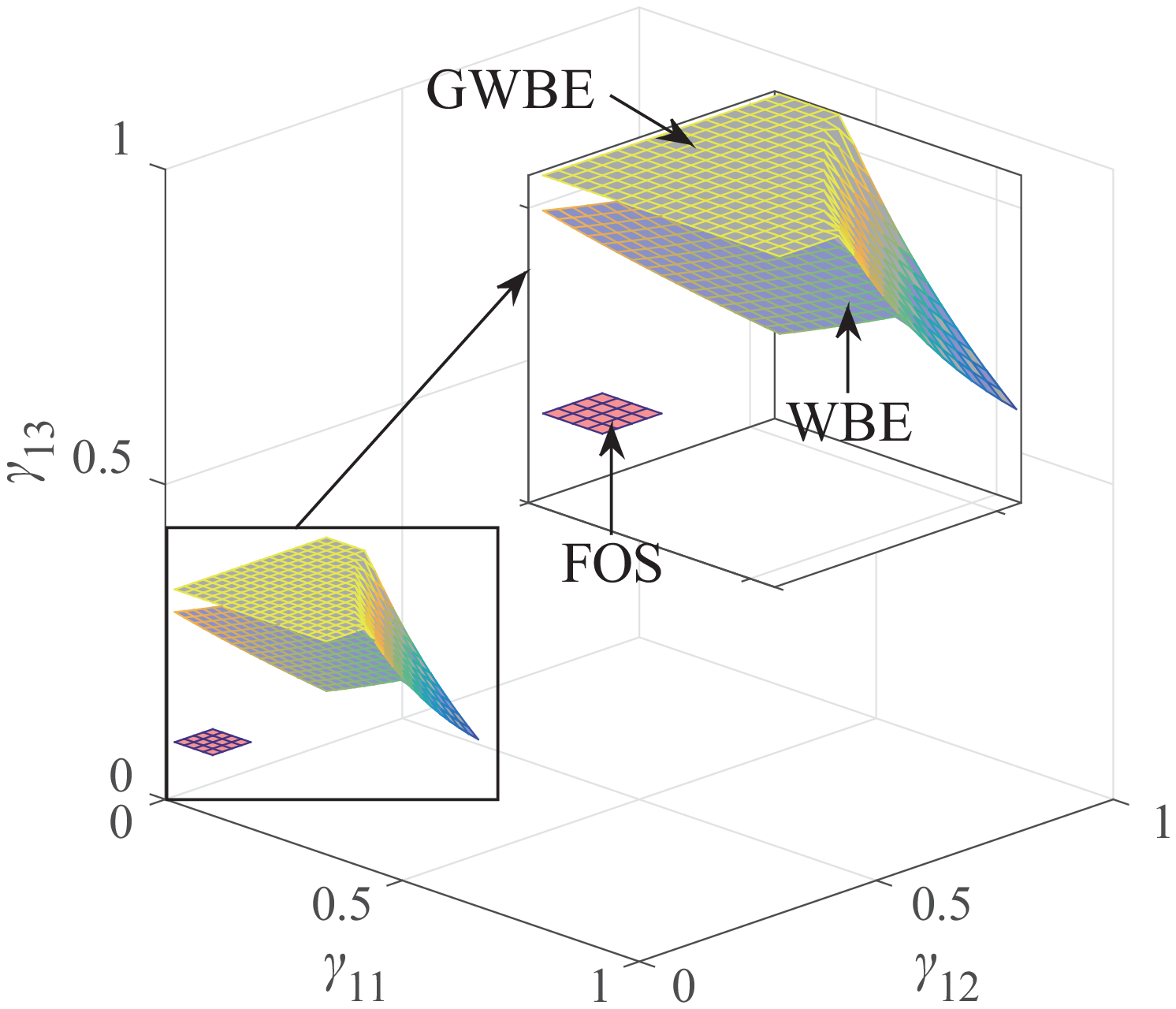}}
\caption{The upper surface boundary of the user capacity regions versus the SINR requirements for the three designs with (a) $L=2$, (b) $L=3$, and (c) $L=4$.}
\label{capacity_region}
\end{minipage}
\end{figure*}

In the WBE design \cite{Welch1974,Tropp2004}, the generated pilot sequences have the same correlation coefficient between each other, which is given by $\rho_{ijlk}=\sqrt{\left(K-\tau\right)/\left(\left(K-1\right)\tau\right)}$, where $\left(i,j\right)\neq\left(l,k\right)$. %As such, this design is different from our design which has distinct correlation coefficients between different pilot sequences.
Given the same correlation coefficient, the parameter $\alpha_{lk}$ in the WBE design is constant for all the users in the network, i.e., $\alpha_{lk}=\alpha$. Accordingly, the transmit power for $\textrm{U}_{lk}$ with the SINR requirement $\gamma_{lk}$ is given by $P_{lk}=\frac{\alpha\gamma_{lk}}{1+\gamma_{lk}}$. Furthermore, the pilot sequence assigned to $\textrm{U}_{lk}$ is identical to that assigned to other $L-1$ users in the network. As such, we denote $\mathcal{U}_{\mathbf{s}_{lk}}^{\textrm{WBE}}$ as the group of users that are assigned the same pilot sequence as $\textrm{U}_{lk}$ and denote $\bar{\mathcal{U}}_{\mathbf{s}_{lk}}^{\textrm{WBE}}$ as the group of users that are assigned different pilot sequences. We highlight that the proposed GWBE design outputs WBE sequences when the SINR requirements of all the users in the network are identical.

\begin{lemma}\label{lemma_wbe}
To satisfy the SINR requirements of all the users in a multi-cell multiuser massive MIMO network using the WBE design, the bound on per-cell user capacity region is given by
\begin{align}\label{WBE}
\sum_{j=1}^{K}\frac{\gamma_{ij}}{1+\gamma_{ij}}
\leq\min\left\{\frac{\tau}{L},\frac{\kappa}{L}+\frac{\left(1-\kappa\right)
\gamma_{i}^{\textrm{MAX}}}{1+\gamma_{i}^{\textrm{MAX}}}\right\},
\end{align}
where $\kappa=\frac{1}{\rho_{ijlk}^{2}}=\frac{\left(K-1\right)\tau}{K-\tau}$.
\begin{IEEEproof}
Please refer to Appendix \ref{lemma_wbe_proof}.
\end{IEEEproof}
\end{lemma}

Using the procedure described in Section \ref{sec:lim_anteana}, we derive the minimum $M$ required by $\textrm{U}_{lk}$ for the WBE design as
\small
\begin{align}\label{antenna_wbe}
M_{lk,\text{WBE}}^{\textrm{MIN}}=\frac{ \sum_{m,n} \beta_{lkm}P_{mn}+\sigma_{w}^{2}}
{\left(\frac{1-\mu}{\alpha\mu}\right)
\left(\sum_{p,q}\Xi_{lkp}^2\beta_{lkp}P_{pq}+ \overline{\kappa}_{lk}\right)},
\end{align}
\normalsize
where $\overline{\kappa}_{lk}=\sum_{r,s} \nicefrac{\Xi_{lkr}^2\beta_{lkr}P_{rs}}{\kappa} -\beta_{lkl}P_{lk}$, $P_{pq}$ is the transmit power for $\textrm{U}_{pq}$, where $\textrm{U}_{pq}\in{}\mathcal{U}_{\mathbf{s}_{pq}}^{\textrm{WBE}}$, and $P_{rs}$ is the transmit power for $\textrm{U}_{rs}$, where $\textrm{U}_{rs}\in\bar{\mathcal{U}}_{\mathbf{s}_{pq}}^{\textrm{WBE}}$. Accordingly, we obtain the minimum $M$ for the network with the WBE design as $M_{\textrm{WBE}}^{\textrm{MIN}}=\max_{1\leq{}l\leq{}L,1\leq{}k\leq{}K}\{M_{lk,\textrm{WBE}}^{\textrm{MIN}}\}$. %The comparison between $M_{\textrm{WBE}}^{\textrm{MIN}}$ and $M_{\textrm{GWBE}}^{\textrm{MIN}}$ will be provided in Section \ref{sec:num_results_fin_M}.

\subsection{Finite Orthogonal Set (FOS) Design}\label{sec:FOS_desc}

In the FOS design \cite{Jose2011,Marzetta2010}, the first $\tau$ users in each cell are assigned the unique $\tau$ orthogonal pilot sequences. Then the remaining $K-\tau$ users in the same cell are repeatedly assigned the same pilot sequences. Given this assignment strategy, the pilot sequence allocated to $\textrm{U}_{lk}$ is used by at least $L-1$ users in the network. We denote  $\mathcal{U}_{\mathbf{s}_{lk}}^{\textrm{FOS}}$ as the group of users that are assigned the same pilot sequence as $\textrm{U}_{lk}$. %It is noteworthy that the correlation coefficient between different pilot sequences is always zero. Therefore, only the users with the same pilot sequence contribute to pilot contamination.
The transmit power for $\textrm{U}_{lk}$ is given by $P_{lk}=\frac{\alpha_{lk}\gamma_{lk}}{1+\gamma_{lk}}$.

\begin{lemma}\label{lemma_fos}
To satisfy the SINR requirements of all the users in a multi-cell multiuser massive MIMO network using the FOS design, the bound on per-cell user capacity region is given by
\begin{align}\label{FOS}
\sum_{i=1}^{K}\left(\frac{\gamma_{ij}}{1+\gamma_{ij}}\right)
\leq\min\left\{\frac{\tau}{L},\frac{1}{L}\right\}.
\end{align}
\begin{IEEEproof}
%We obtain \eqref{FOS} by following the procedure outlined in
Please refer to Appendix \ref{lemma_wbe_proof}.
\end{IEEEproof}
\end{lemma}

We find that the minimum $M$ for the network with the FOS design is given by $M_{\textrm{FOS}}^{\textrm{MIN}}=\max_{1\leq{}l\leq{}L,1\leq{}k\leq{}K}\{M_{lk,\textrm{FOS}}^{\textrm{MIN}}\}$, where $M_{lk,\textrm{FOS}}^{\textrm{MIN}}$ is obtained as
\begin{align}\label{antenna_fos}
M_{lk,\text{FOS}}^{\textrm{MIN}}=\frac{\sum_{m,n}\beta_{lkm}P_{mn} +\sigma_{w}^{2}}{
\left(\frac{1-\mu}{\mu}\right)\left(\sum_{i,j} \frac{\Xi_{lki}^2\beta_{lki}P_{ij}}{\alpha_{ij}}-\beta_{lkl}P_{lk}\right)}
\end{align}
with $P_{ij}$ denoting the transmit power for $\textrm{U}_{ij}$ and $\textrm{U}_{ij}\in{}\mathcal{U}_{\mathbf{s}_{ij}}^{\textrm{FOS}}$. We will compare $M_{\textrm{FOS}}^{\textrm{MIN}}$ with $M_{\textrm{GWBE}}^{\textrm{MIN}}$ in Section \ref{sec:num_results_fin_M}.

\section{Numerical Results}\label{sec:num_results}

In this section, we provide numerical results to demonstrate the advantages of using the proposed user capacity-achieving GWBE pilot sequence design over existing pilot sequence designs, namely, the WBE and FOS designs.

\subsection{Performance Comparison with Infinite Antennas}\label{sec:num_results_inf_M}

In this subsection, we assume that $M$ is infinite and $\tau=3$. We compare the user capacity region of the three designs and examine the impact of increasing $K$, varying SINR requirements, and increasing $L$ on the network performance. %Throughout this subsection, we consider %that the length of pilot sequence is
In this subsection, the results for the GWBE, WBE, and FOS designs are generated by using \eqref{BW}, \eqref{WBE}, and \eqref{FOS}, respectively.

We first consider the multiuser massive MIMO network with $L=2$, $L=3$, and $L=4$, where each cell has $K=4$ users. The SINR requirements in each cell are chosen as $\mathbf{\gamma}_{i}=\left[\gamma_{i1},\gamma_{i2},\gamma_{i3},0.20\right]$, where $i\in\left\{2,3,4\right\}$. Fig.~\ref{capacity_region} depicts the upper surface boundary of the per-cell user capacity region of the three designs. We observe that the proposed GWBE pilot design achieves a larger per-cell user capacity region than the WBE and FOS designs. Importantly, this observation is valid for different values of $L$. This indicates that the proposed GWBE design is more likely to fulfill a diverse set of SINR requirements of the users. Numerically,
the proposed GWBE design has 19.2\% and 97.4\% larger per-cell user capacity region than the WBE and FOS designs, respectively, for a four cell massive MIMO network.

\begin{figure}[!t]
\centering
\includegraphics[width=21pc]{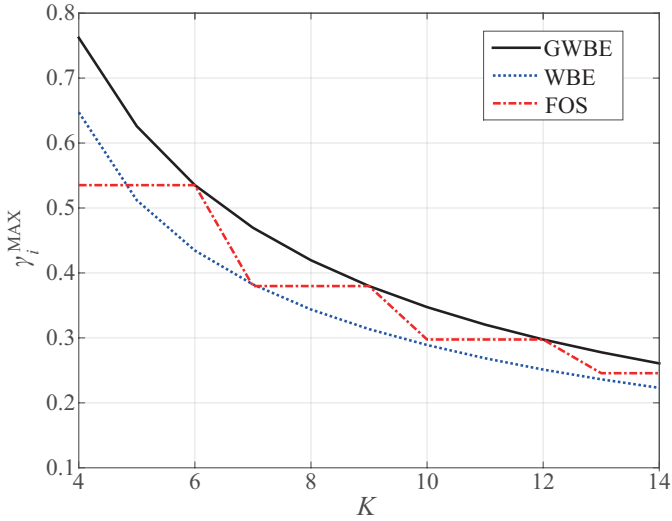}
\caption{The per-cell maximum permitted SINR requirement versus the number of per-cell users for the three designs with $L=2$ and $\tau=3$.}
\label{num_user}
\end{figure}
\begin{figure}[!t]
\centering
\includegraphics[width=21pc]{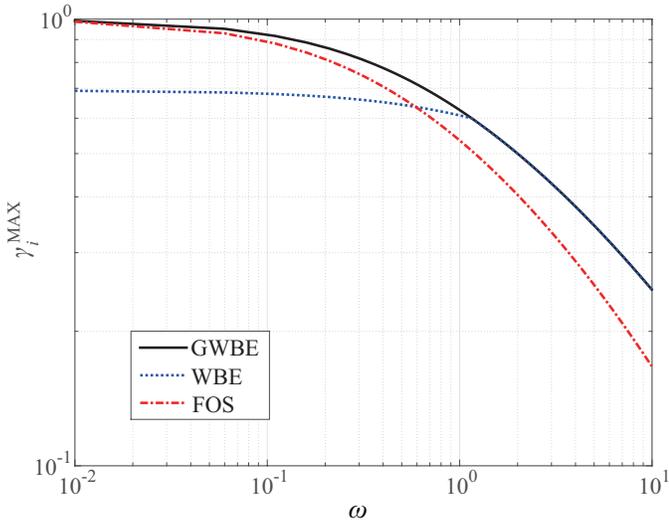}
\caption{The per-cell maximum permitted SINR requirement versus $\omega$ for the three designs with $L=2$ and $K=5$.}
\label{max_SNR}
\end{figure}

We next examine the impact of increasing $K$ on the maximum permitted SINR requirement in a cell, $\gamma_{i}^{\textrm{MAX}}$. Fig.~\ref{num_user} depicts $\gamma_{i}^{\textrm{MAX}}$ versus $K$ for the three designs. In this figure, we consider $L=2$ and the same SINR requirements for the users in two cells given by $\pmb{\gamma}_{i}=\left[\gamma_{i1},\dotsc,\gamma_{i3}, \gamma_{i4},\dotsc,\gamma_{iK}\right]$, where $i\in\{1,2\}$, $\gamma_{i1}=\gamma_{i2}=\gamma_{i3}=\gamma$, and $\gamma_{i4}=\dotsc=\gamma_{iK}={\gamma}/{3}$. As such, we have $\gamma_{i}^{\textrm{MAX}}=\gamma$. We first see that increasing $K$ decreases $\gamma_{i}^{\textrm{MAX}}$. For example, when $K$ increases from $4$ to $14$, $\gamma_{i}^{\textrm{MAX}}$ of our GWBE design decreases from 0.76 to 0.26. This is due to the fact that increasing $K$ causes more interference and thus lowers the maximum permitted SINR. Second, we see that the FOS design behaves differently from other designs as $K$ increases. For example, when $K$ increases from 4 to 6 and from 7 to 9, $\gamma_{i}^{\textrm{MAX}}$ remains constant. This can be explained by the pilot sequence assignment strategy in the FOS design. Since only $\tau$ orthogonal pilot sequences are available in the FOS design, $\gamma_{i}^{\textrm{MAX}}$ decreases only when $K$ exceeds the exact multiple of $\iota\tau$, where $\iota=\textrm{card}\left(\mathcal{U}_{\mathbf{s}_{lk}}^{\textrm{FOS}}\right)$. For instance, $\gamma_{i}^{\textrm{MAX}}$ decreases only when $K>6$ and $K>9$, where $\tau=3$ and $\textrm{card}\left(\mathcal{U}_{\mathbf{s}_{lk}}^{\textrm{FOS}}\right)=2$.

We now evaluate the impact of increasing the individual SINR requirement on $\gamma_{i}^{\textrm{MAX}}$. In this evaluation, we consider $L=2$, $K=5$, and the same SINR requirements of the users in the two cells given by $\pmb{\gamma}_{i}=\left[\gamma,\gamma,\gamma,\omega\gamma,\omega\gamma \right]$, where $i\in\left\{1,2\right\}$ and $10^{-2}\leq\omega\leq10^{1}$. Fig.~\ref{max_SNR} depicts $\gamma_{i}^{\textrm{MAX}}$ versus $\omega$ for the three designs. We first find that our GWBE design outperforms the FOS design across the whole range of $\omega$, and outperforms the WBE design in the low and medium $\omega$ regime. When $\omega>10^{0}$, our GWBE design exhibits the same performance as the WBE design. This is due to the fact that the upper surface boundary of the per-cell user capacity region of the GWBE and WBE designs overlap in this regime. Consequently, the performance of two designs is equivalent to each other. We also find that the FOS design outperforms the WBE design in the low $\omega$ regime, since the FOS design incurs less interference than the WBE design in this regime.

\begin{figure}[!t]
\centering
\includegraphics[height=2.8in,width=3.5in]{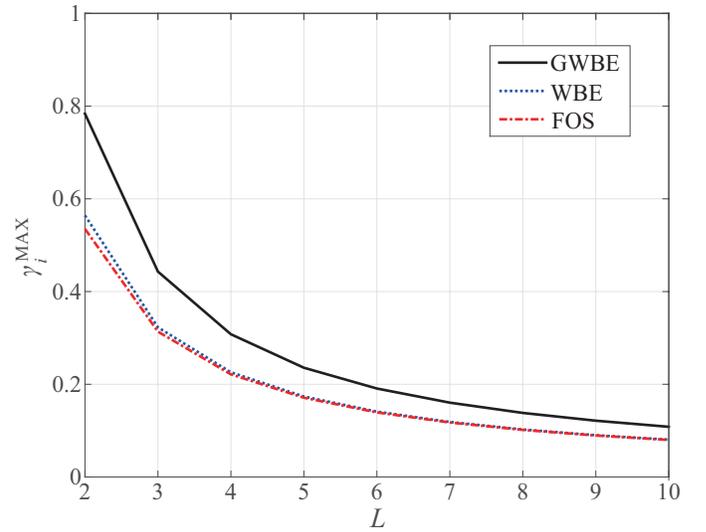}
\caption{The per-cell maximum permitted SINR requirement versus the number of cells for the three designs with $K=5$.}
\label{num_cells}
\end{figure}

Finally, we focus on the impact of increasing $L$ on $\gamma_{i}^{\textrm{MAX}}$. In particular, we consider an $L$-cell network with $K=5$ and the SINR requirements given by $\pmb{\gamma}_{i}=\left[\gamma,\gamma,\gamma/3,\gamma/3,\gamma/3\right]$, where $i\in\left\{1,\dotsc,L\right\}$. Fig. \ref{num_cells} depicts $\gamma_{i}^{\textrm{MAX}}$ versus $L$ for the three designs. We first observe that increasing $L$ decreases $\gamma_{i}^{\textrm{MAX}}$. For example, when $L$ increases from 2 to 10, $\gamma_{i}^{\textrm{MAX}}$ of our GWBE design decreases from 0.78 to 0.11. This can be explained by the decrease in the per-cell user capacity region, as depicted in Fig. \ref{capacity_region}. Second, we observe that the our GWBE design provides higher $\gamma_{i}^{\textrm{MAX}}$ relative to the WBE and FOS designs, resulted from its larger user capacity region.

\subsection{Performance Comparison with Finite Antennas}\label{sec:num_results_fin_M}
\begin{table*}[!t]
\centering
\begin{minipage}{\textwidth}
\renewcommand{\arraystretch}{1.2}
\caption{Simulation Parameters and Pilot Sequences}
\label{table_simulation}
\centering
\begin{tabular}{|c|c|c|c|}
\hline
& \bfseries GWBE & \bfseries WBE & \bfseries FOS \\
\hline\hline
Common Parameters & \multicolumn{3}{c|}{$\sigma_{z}^{2}=\sigma_{w}^{2}=1$, $L=2$, $\tau=3$, $K=4$, $K_{\textrm{tot}}=8$, $\pmb{\gamma}_{1}=\left[0.91, 0.74, 0.64, 0.23\right], \pmb{\gamma}_{2}=\left[0.94, 0.82, 0.45, 0.10\right]$}\\
& \multicolumn{3}{c|}{$\beta_{lkm}=1$, where $l=m$ and $\beta_{lkm}=0.9$, where $l\neq m$}  \\
\hline
Transmit Power & ${\alpha_{lk}\hat{\gamma}_{lk}}/\left({1+\hat{\gamma}_{lk}}\right)$ ${}^{\ast}$ & ${\alpha_{lk}\gamma_{lk}}/\left({1+\gamma_{lk}}\right)$ & ${\alpha_{lk}\gamma_{lk}}/\left({1+\gamma_{lk}}\right)$ \\
\hline
Pilot Sequences & $\mathbf{S}_{1} =
\begin{bmatrix}
 1 &  -0.0845 &  -0.1075  &  0.2574 \\
 0 &   0.9964  &  -0.2202  &  0.5274\\
 0    &     0 &   0.9695 &   0.8097
 \end{bmatrix}$ & $\mathbf{S}_{1} =
\begin{bmatrix}
 1 & -0.3333 &  -0.3333 &   0.3333 \\
 0   & 0.9428  & -0.4714  &  0.4714\\
 0    &     0  &   0.8165  &  0.8165
 \end{bmatrix}$ & $\mathbf{S}_{1} =
\begin{bmatrix}
 1  & 0 & 0 & 1 \\
 0   & 1  & 0  &  0\\
 0    &  0  &   1  &  0
 \end{bmatrix}$\\

 & $\mathbf{S}_{2} =
\begin{bmatrix}
 1  & -0.0500  & -0.1182 &   0.1867 \\
 0   & 0.9988 &  -0.2186  &  0.3453\\
 0    &  0  &  0.9686  & 0.9197
 \end{bmatrix}$ & $\mathbf{S}_{2} = \mathbf{S}_{1}$ & $\mathbf{S}_{2} = \mathbf{S}_{1}$ \\
 \hline
\multicolumn{4}{l}{{${}^\ast$}$\hat{\pmb{\gamma}}_{1}=\left[0.92, 0.75, 0.65, 0.24\right]$,  $\hat{\pmb{\gamma}}_{2}=\left[0.95, 0.85, 0.50, 0.29\right]$}%\\
\end{tabular}
\end{minipage}
\end{table*}

\begin{figure*}[!t]
\centering
\begin{minipage}{\textwidth}
\subfloat[$\theta_{11,M}$ and $\theta_{14,M}$]{{\includegraphics[width=21pc]{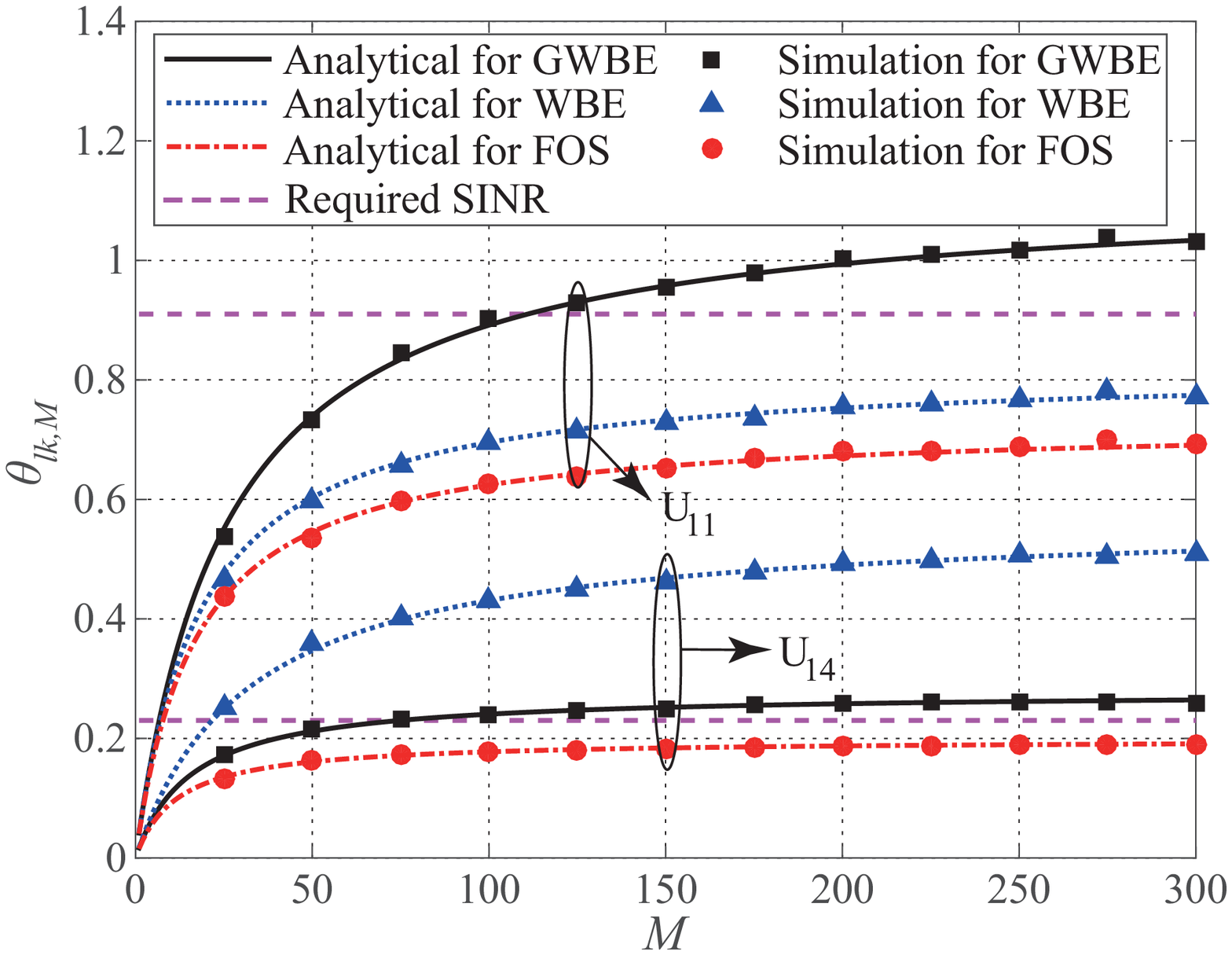} }}
\subfloat[$\theta_{22,M}$ and $\theta_{23,M}$]{{\includegraphics[width=21pc]{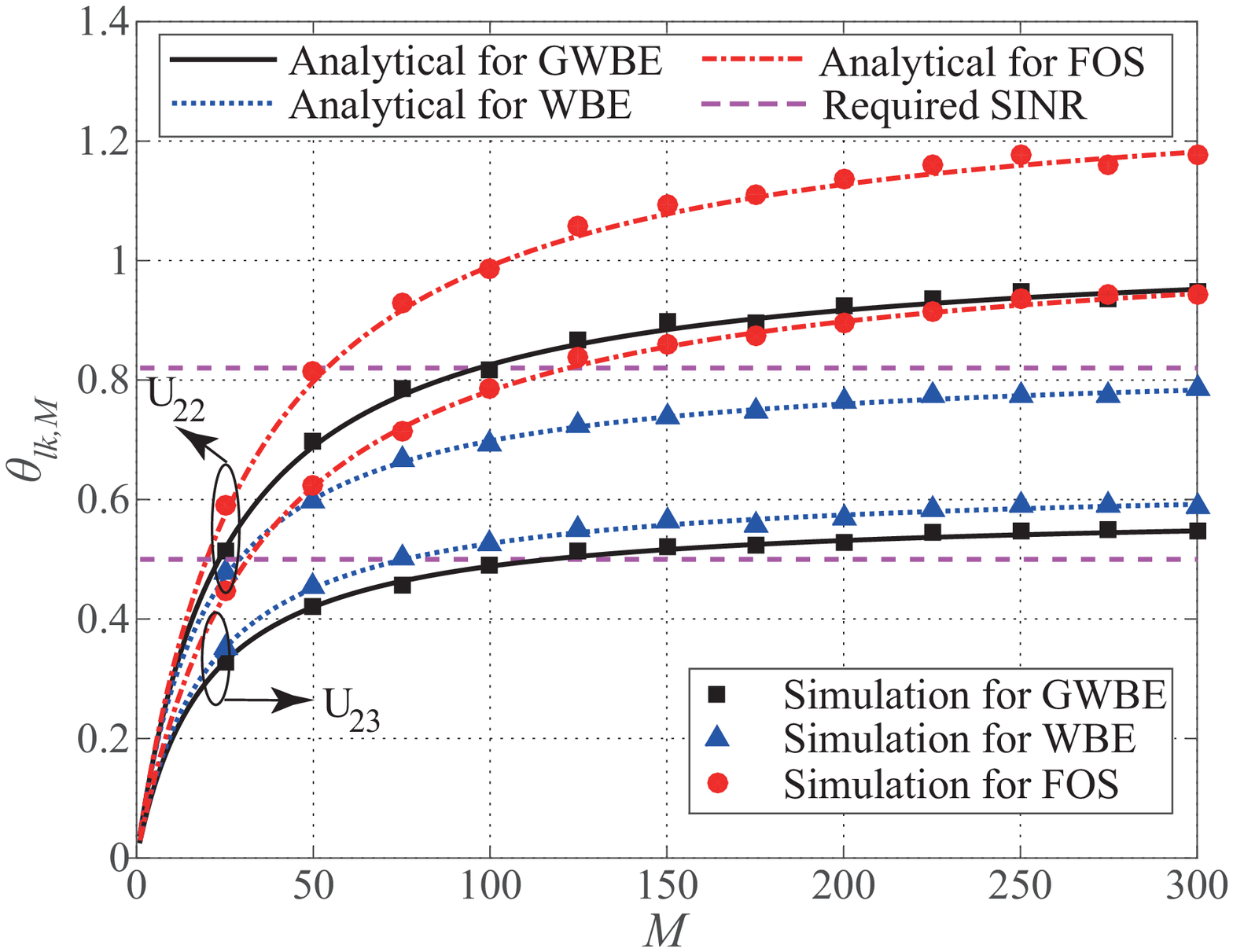} }}
\caption{The achievable SINR versus the number of antennas for the three designs.}
\label{m_antenna_SINR}
\end{minipage}
\end{figure*}
In this subsection, we assume that $M$ is finite and compare the performance achieved by the three designs.
Throughout this subsection, we consider $L=2$, $\tau=3$, and $K=4$. We also consider that the SINR requirements in the two cells are given by $\pmb{\gamma}_{1}=\left[0.91, 0.74, 0.64, 0.23\right]$ and $\pmb{\gamma}_{2}=\left[0.94, 0.82, 0.45, 0.20\right]$. We note that the considered SINR requirements remain within the user capacity region of our GWBE design but lie outside the user capacity region of the WBE and FOS designs. This implies that our GWBE design supports a more diverse range of SINR requirements than other designs. In addition, we note that an important criterion in designing the user capacity-achieving GWBE pilot sequences is that $\hat{\gamma}_{lk}>\gamma_{lk}$ needs to be chosen to satisfy \eqref{BW_all} with equality, as clarified in Section \ref{sec:design}. As such, we choose $\pmb{\hat{\gamma}}_{1}=\left[0.92, 0.75, 0.65, 0.24\right]$ and $\pmb{\hat{\gamma}}_{2}=\left[0.95, 0.85, 0.50, 0.29\right]$ for the two cells. Furthermore, we assume that $p_{ij}\beta_{ijl} = 0.9$, where $i \neq l$. For clarity, the simulation parameters and the pilot sequences for the three designs are summarized in Table \ref{table_simulation}. Based on the runtime, we first note that the proposed GWBE \textbf{Algorithm~\ref{algo1}} depicts polynomial computational complexity with the increase in $L$ and $K$. We also note that the complexity remains constant with the increase in $\tau$. Consequently, the overall computational complexity of the proposed algorithm is given as $\mathcal{O}\left(K^c\right)$ where $c$ ranges from $1.41$ to $1.88$ in our simulations. The WBE design has the same computational complexity as our GWBE design, but achieves worse performance. We further note that the FOS design has a lower computational complexity than our proposed GWBE design. However, we highlight that the performance benefit of using the proposed GWBE design outweighs the complexity benefit offered by the FOS design as the FOS design is unable to meet the SINR requirements of all the users in the network.

Fig.~\ref{m_antenna_SINR} depicts $\theta_{lk,M}$ versus $M$ for the three designs. In this figure, we only consider $\theta_{lk,M}$ of four users, namely, $\textrm{U}_{11}$, $\textrm{U}_{14}$, $\textrm{U}_{22}$, and $\textrm{U}_{23}$, to avoid visual cluttering\footnote{We clarify that the behavior  of the unconsidered users, i.e., $\textrm{U}_{12}$, $\textrm{U}_{13}$, $\textrm{U}_{21}$, and $\textrm{U}_{24}$, is similar to that of the four users considered in Fig.~\ref{m_antenna_SINR}.}. The analytical curves for the three designs are generated from \eqref{SINR}, while the Monte Carlo simulation points for the three designs are obtained by averaging \eqref{long_exp} over 1,000 channel realizations. It is clearly seen that the Monte Carlo simulation points precisely agree with the analytical curves, which demonstrates the accuracy of \eqref{SINR}. We first see from Fig. \ref{m_antenna_SINR} that our GWBE design is the only design that satisfies the SINR requirements of all the users in the network. If the WBE design or the FOS design is adopted, the SINR requirements of only a few users are satisfied. Second, we see that our GWBE design does not always provide the highest achievable SINR for each user. This can be explained by $\alpha_{lk}$ for $\textrm{U}_{lk}$ of the three designs as follows: As indicated by \eqref{SINR}, an increase in $\alpha_{lk}$ gives a lower achievable SINR. The value of $\alpha_{lk}$ of the eight users for the three designs are summarized in Table \ref{correlation coefficient}. Using Fig.~\ref{m_antenna_SINR} together with Table \ref{correlation coefficient}, we find that the highest achievable SINR is attained for the lowest $\alpha_{lk}$ except for $l=2,k=4$. Furthermore, we point out that the WBE and FOS designs provide a higher achievable SINR for some users than our GWBE design. However, the higher achievable SINR is provided at the expense of degrading the achievable SINR of other users. As such, the user with the degraded achievable SINR cannot satisfy the pre-defined SINR requirements even as $M\rightarrow\infty$. Therefore, it is worth highlighting that the practical advantage of our GWBE design lies in its ability of fulfilling the SINR requirements of all the users in the network. We highlight that the proposed pilot design controls the correlation between different pilot sequences and the downlink power allocation keeps the interference at an acceptable level. Therefore, the proposed pilot design and downlink power allocation scheme mitigate pilot contamination in the massive MIMO network.
\begin{table}[!t]
\renewcommand{\arraystretch}{1.1}
\caption{Summary of the Value of $\alpha_{lk}$}
\label{correlation coefficient}
\centering
\begin{tabular}{|c|c|c|c|c|c|c|}
\hline
& \multicolumn{2}{|c|}{\bfseries GWBE } & \multicolumn{2}{|c|}{\bfseries WBE } & \multicolumn{2}{|c|}{\bfseries FOS } \\
\hline
$\alpha_{lk}$  & \textit{l}=1 & \textit{l}=2 & \textit{l}=1 & \textit{l}=2 & \textit{l}=1 & \textit{l}=2 \\
\hline
\textit{k}=1 & ${3.03}^{\ast\circ}$ & ${3.03}^{\ast\circ}$ & 3.53 & 3.53 & 4.80 & 4.80 \\
\hline
\textit{k}=2 & ${3.35}^{\circ}$ & ${3.34}^{\circ}$ & ${3.53}^{\circ}$ & ${3.53}^{\circ}$ & ${2.90}^{\ast\circ}$ & ${2.90}^{\ast\circ}$ \\
\hline
\textit{k}=3 & ${3.99}^{\circ}$ & ${4.01}^{\circ}$ & ${3.53}^{\circ}$ & ${3.53}^{\circ}$ & ${2.90}^{\ast\circ}$ & ${2.90}^{\ast\circ}$ \\
\hline
\textit{k}=4 & ${4.25}^{\circ}$ & ${4.33}^{\circ}$ & ${3.53}^{\ast\circ}$ & ${3.53}^{\ast\circ}$ & 4.80 & 4.80 \\
\hline
\multicolumn{7}{l}{${}^{\ast}$ indicates the highest achievable SINR; ${}^{\circ}$ indicates that $\theta_{lk} \geq \gamma_{lk}$.}
\end{tabular}
\end{table}
\begin{figure}[!t]
\centering
\includegraphics[height=2.8in,width=3.5in]{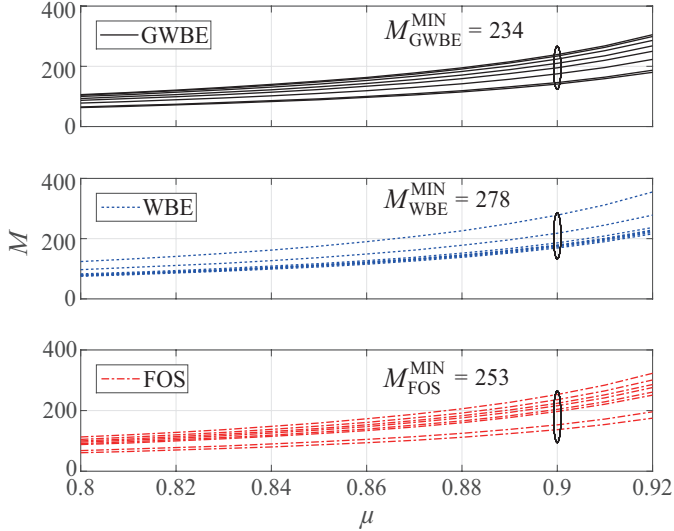}
\caption{The number of antennas required at BSs versus the performance satisfaction index for the three designs.}\label{num_antenna_percent}
\end{figure}

Finally, we find the minimum number of antennas at BSs to achieve the satisfactory network performance. In this evaluation, we consider the same parameters as given in Table \ref{table_simulation}. Fig. \ref{num_antenna_percent} depicts the minimum number of antennas required by the eight users to achieve the SINR requirement given by $\mu\theta_{lk,\infty}$ for the three designs. The minimum number of antennas clarified in the figure, i.e., $M_{\textrm{GWBE}}^{\textrm{MIN}}=234$, $M_{\textrm{WBE}}^{\textrm{MIN}}=278$, and $M_{\textrm{FOS}}^{\textrm{MIN}}=253$, are determined by setting $\mu=0.9$ in \eqref{antenna_gwbe}, \eqref{antenna_wbe}, and \eqref{antenna_fos}, respectively. Comparing $M_{\textrm{GWBE}}^{\textrm{MIN}}$ with $M_{\textrm{WBE}}^{\textrm{MIN}}$ and $M_{\textrm{FOS}}^{\textrm{MIN}}$, we find that a key benefit of our GWBE design is that it requires 15.8\% and 7.5\% less antennas than the WBE and FOS design, respectively, to achieve similar performance. This benefit becomes particularly compelling when considering the fact that the GWBE design achieves the SINR requirements of all the users in the network.

\section{Conclusion}\label{conclusion}

We proposed a novel GWBE pilot sequence design for multi-cell multiuser massive MIMO networks. We first derived closed-form expressions for the user capacity and the user capacity region of the network. Based on them, we developed a new algorithm to generate user capacity-achieving pilot sequences which always fulfill the SINR requirements of all users throughout the network. We further conducted analytical and numerical performance comparison of our proposed GWBE design with the existing WBE and FOS designs. The comparison demonstrated that our GWBE design achieves a larger user capacity region and supports a more diverse range of SINR requirements than the WBE and FOS designs. In addition, we confirmed that our GWBE design needs a lower number of antennas at BSs than the WBE and FOS designs to meet the predefined SINR requirements. Our results offer a set of practically valuable guidelines for next-generation wireless infrastructure providers to efficiently design pilot sequences such that superior performance is achieved without utilizing unnecessary antennas.

\appendices

\section{Proof of Lemma \ref{lemma_red}}\label{SINR_proof}
In this appendix, we detail the derivation of a channel-independent expression for the achievable SINR, given by \eqref{SINR}, based on the channel-dependent expression for the achievable SINR given by \eqref{long_exp}. First, we calculate the mathematical expectation, $\mathbb{E}\left[\left|\mathbf{g}_{lk}^{mn}\right|^{2}\right]$, in the denominator of \eqref{long_exp}. Using \eqref{precoding_vec}, we rewrite $\mathbb{E}\left[\left|\mathbf{g}_{lk}^{mn}\right|^{2}\right]$ as
\begin{align}\label{precoding_all}
\mathbb{E}\left[\left|\mathbf{g}_{lk}^{mn}\right|^{2}\right]=
\mathbb{E}\left[\left|\mathbf{h}_{lkm}^{H}\mathbf{t}_{mn}\right|^{2}\right]=
\mathbb{E}\left[\left|\frac{\mathbf{h}_{lkm}^{H}\hat{\mathbf{h}}_{mnm}} {\sqrt{M\alpha_{mn}}}\right|^{2}\right].
\end{align}
Using the LS channel estimate given by \eqref{channel_estimate_2} and defining $\mathbf{\breve{g}}_{ijm} \triangleq \Xi_{ijm}\mathbf{h}_{ijm}$, $\sum_{i=1}^{L}\sum_{j=1}^{K}\triangleq \sum_{i,j}$, we expand $\left|\mathbf{h}_{lk}^{H}\hat{\mathbf{h}}_{mnm}\right|^2$ in \eqref{precoding_all} as
\begin{align}\label{interme-appendix}
\left|\mathbf{h}_{lkm}^{H}\hat{\mathbf{h}}_{mnm}\right|^2&=\sum_{p,q}\sum_{r,s}\rho_{pqmn}\rho_{rsmn}\mathbf{\breve{g}}_{pqm}^{H}
\mathbf{h}_{lkm}\mathbf{h}_{lkm}^{H}\mathbf{\breve{g}}_{rsm} \notag \\ &+ \sum_{p,q}\rho_{pqmn}\mathbf{\breve{g}}_{pqm}^{H}\mathbf{h}_{lkm}
\mathbf{h}_{lkm}^{H}\mathbf{S}_{mn}^{T}\mathbf{z}_{m} \notag \\
&+\mathbf{z}_{m}^{H}\mathbf{S}_{mn}\sum_{r,s}\rho_{rsmn}
\mathbf{h}_{lkm}\mathbf{h}_{lkm}^{H}\mathbf{\breve{g}}_{rsm} \notag \\ &+ \mathbf{z}_{m}^{H}\mathbf{S}_{mn}\mathbf{h}_{lkm} \mathbf{h}_{lkm}^{H}
\mathbf{S}_{mn}^{T}\mathbf{z}_{m}.
\end{align}
Defining $\sum_{p,q}\sum_{r,s}\triangleq\sum_{\substack{p,q\\r,s}}$, we calculate the expectation of \eqref{interme-appendix} as
\begin{align}\label{interme-appendix2}
\mathbb{E}\left[\left|\mathbf{h}_{lkm}^{H}\hat{\mathbf{h}}_{mnm}\right|^2 \right]
&=\mathbb{E}\left[\sum_{\substack{p,q\\r,s}}\rho_{pqmn}\rho_{rsmn} \mathbf{\breve{g}}_{pqm}^{H}
\mathbf{h}_{lkm}\mathbf{h}_{lkm}^{H}\mathbf{\breve{g}}_{rsm}\right] \notag \\&+0 + 0 +  \mathbb{E}\left[\mathbf{z}_{m}^{H}\mathbf{S}_{mn}\mathbf{h}_{lkm} \mathbf{h}_{lkm}^{H}
\mathbf{S}_{mn}^{T}\mathbf{z}_{m} \right],\notag\\
&= \rho_{lkmn}^2\mathbb{E}\left[\mathbf{\breve{g}}_{lkm}^{H} \mathbf{h}_{lkm} \mathbf{h}_{lkm}^{H} \mathbf{\breve{g}}_{lkm}\right] \notag \\&+\sum_{\substack{p=r\ne l\\q=s \ne k}}\rho_{pqmn}^2\mathbb{E}\left[\textrm{tr}\left(\mathbf{\breve{g}}_{pqm}^{H}\mathbf{h}_{lkm} \mathbf{h}_{lkm}^{H}\mathbf{\breve{g}}_{pqm}\right)\right]\notag\\&+\mathbb{E}\left[\textrm{tr}\left(\mathbf{z}_{m}^{H}\mathbf{S}_{mn} \mathbf{h}_{lkm}\mathbf{h}_{lkm}^{H}\mathbf{S}_{mn}^{T}\mathbf{z}_{m} \right)\right].
\end{align}
We observe that $\mathbf{h}_{lkm}^{H}\mathbf{h}_{lkm}$ in \eqref{interme-appendix2} has a gamma distribution with the shape parameter $M$ and the scale parameter $1$ given by $\Gamma\left(M,1\right)$. Therefore, we find that $\mathbb{E}\left[\mathbf{\breve{g}}_{lkm}^{H}\mathbf{h}_{lkm} \mathbf{h}_{lkm}^{H} \mathbf{\breve{g}}_{lkm}\right]$ in \eqref{interme-appendix2} is $\Xi_{lkm}^{2} M\left(M+1\right)$. This simplifies \eqref{interme-appendix2} as
\begin{align}\label{interme-appendix3}
\mathbb{E}\left[\left|\mathbf{h}_{lkm}^{H}\hat{\mathbf{h}}_{mnm}\right|^2\right]
&=\rho_{lkmn}^2 \Xi_{lkm}^{2} M\left(M+1\right) \notag\\ &+\sum_{\substack{p=r\ne l\\q=s \ne k}} \rho_{pqmn}^2 \mathbb{E}\left[\textrm{tr}\left(\mathbf{h}_{lkm}\mathbf{h}_{lkm}^{H} \mathbf{\breve{g}}_{pqm}\mathbf{\breve{g}}_{pqm}^{H}\right)\right] \notag \\&+\mathbb{E}\left[\textrm{tr}\left(\mathbf{h}_{lkm} \mathbf{h}_{lkm}^{H}\mathbf{S}_{mn}^{T}\mathbf{z}_{m}^{H}\mathbf{z}_{m}^{H}\mathbf{S}_{mn} \right)\right],\notag\\
&=\rho_{lkmn}^2\Xi_{lkm}^{2}\left(M^2+M\right) \notag \\&+ M\sum_{p=r \ne l}^{L}\sum_{q=s \ne k}^{K} \rho_{pqmn}^2\Xi_{pqm}^2 + M\sigma_{z}^2,\notag\\
&=M^2\rho_{lkmn}^2\Xi_{lkm}^{2}+M\alpha_{mn}.
\end{align}
Substituting \eqref{interme-appendix3} into \eqref{precoding_all}, we obtain the expression $\mathbb{E}\left[\left|\mathbf{g}_{lk}^{mn}\right|^2\right]=\left(M\rho_{lkmn}^2\Xi_{lkm}^{2}+\alpha_{mn}\right)/\alpha_{mn}$. We observe that the variance of the gamma distributed $\mathbf{g}_{lk}$ is given by $\text{var}\left[\mathbf{h}_{lkl}^{H} \hat{{\mathbf{t}}}_{lk}\right] = 1$.
\begin{table*}[!t]
\centering
\begin{minipage}{\textwidth}
\renewcommand{\arraystretch}{1.2}
\caption{Matrix Definitions}
\label{matrix_definitions}
\centering
\begin{tabular}{ccc}
$\mathbf{S}_{l} =
\begin{bmatrix}
\mathbf{s}_{l1} & \mathbf{s}_{l2} &  \dotsc &   \mathbf{s}_{lK}
\end{bmatrix}$, &

 $\mathbf{P}_{l} =  \text{diag}\left\{\mathbf{p}_{l}\right\} =
\begin{bmatrix}
P_{l1}  & 0         &  \dotsc &\dotsc &0\\
0       & P_{l2} &  0 & \dotsc &0 \\
\vdots  & 0 &  \ddots & 0 &0  \\
\vdots  & \vdots &  0 & \ddots &0 \\
0   & 0 &  0 & 0 & P_{lK} \\
 \end{bmatrix}$,  &

$\mathbf{A}_{l} =  \text{diag}\left\{\mathbf{a}_{l}\right\} =
\begin{bmatrix}
\frac{1}{\alpha_{l1}} & 0         &  \dotsc &\dotsc &0\\
0          & \frac{1}{\alpha_{l2}} &  0 & \dotsc &0 \\
\vdots   & 0 &  \ddots & 0 &0  \\
\vdots   & \vdots &  0 & \ddots &0 \\
0   & 0 &  0 & 0 & \frac{1}{\alpha_{lK}} \\
 \end{bmatrix}$  \\
 \\\hline
\end{tabular}
\end{minipage}
\end{table*}

Finally, substituting \eqref{interme-appendix3} and value of $\text{var}\left[\mathbf{h}_{lkl}^{H} \hat{{\mathbf{t}}}_{lk}\right]$ into \eqref{long_exp}, $\theta_{lk,M}$ is derived as
\small
\begin{align}\label{interme-appendix52}
\theta_{lk,M} &= \frac{\left(\frac{M}{\sqrt{M\alpha_{lk}}}\right)^2\beta_{lkl}P_{lk}} {\beta_{lkl}P_{lk} + \sum\limits_{m,n \neq l,k} \left( \frac{M\rho_{lkmn}^2\Xi_{lkm}^{2}+\alpha_{mn}}{\alpha_{mn}} \right)\beta_{lkm}P_{mn} + \sigma_{w}^2}, \notag \\
&=\frac{\beta_{lkl}P_{lk}}{\frac{\alpha_{lk}}{M}\left[{\sum\limits_{m,n \neq l,k} \left( \frac{M\rho_{lkmn}^2\Xi_{lkm}^{2}}{\alpha_{mn}} \right)\beta_{lkm}P_{mn} + \bar{P}_{lk}}\right]},
\end{align}
\normalsize
where $\bar{P}_{lk} = \sum_{m,n}\beta_{lkm}P_{mn} + \sigma_{w}^2$.
Simplifying \eqref{interme-appendix52} we obtain the desired result in \eqref{SINR}.

\section{Proof of Proposition \ref{prop_K_tot}}\label{prop1_proof}
In this appendix, we determine the user capacity of the multi-cell multiuser massive MIMO network. Based on \eqref{SINR_inf} and the rule of uplink power control given by $p_{lk}\beta_{lkm}\leq 1$, we obtain
\begin{align}\label{assumption_pow}
  \theta_{lk,\infty} \geq \hat{\theta}_{lk,\infty} &= \frac{P_{lk}} { \alpha_{lk}\sum_{m,n}\frac{\rho_{lkmn}^{2}P_{mn}}{\alpha_{mn}} - P_{lk}}, \notag\\&= \frac{P_{lk}}{\alpha_{lk}\textrm{tr}\left(\mathbf{s}_{lk}^T \mathbf{S}\mathbf{P}\mathbf{A} \mathbf{S}^{T}\mathbf{s}_{lk}\right)},
\end{align}
where $\mathbf{S}$, $\mathbf{P}$, and $\mathbf{A}$ are block matrices given by $\mathbf{S}=\left[\mathbf{S}_{1},\dotsc, \mathbf{S}_{l},\dotsc, \mathbf{S}_{L}\right]$, $\mathbf{P}=\text{diag}\left[\mathbf{P}_{1},\dotsc, \mathbf{P}_{l},\dotsc, \mathbf{P}_{L}\right]$, and $\mathbf{A}=\text{diag}\left[\mathbf{A}_{1},\dotsc,\mathbf{A}_{l},\dotsc, \mathbf{A}_{L}\right]$, respectively.
The matrices $\mathbf{S}_{l}$, $\mathbf{P}_{l}$, and $\mathbf{A}_{l}$ are given by Table \ref{matrix_definitions}.

We next obtain the following expression
\begin{align}\label{interme-appendix6}
\sum_{i,j}\frac{1+\hat{\theta}_{ij,\infty}}{\hat{\theta}_{ij,\infty}}%&=
%\sum_{i,j}\left[1+\frac{\alpha_{ij}\textrm{tr}\left(\mathbf{s}_{ij} \mathbf{S}\mathbf{P}\mathbf{A} \mathbf{S}^{T}\mathbf{s}_{ij}\right)}{P_{ij}} \right],\notag\\
=\textrm{tr}\left(\mathbf{P}^{-1}\mathbf{A}^{-1}\mathbf{S}^{T}\mathbf{S} \mathbf{P}\mathbf{A}\mathbf{S}^{T}\mathbf{S}\right).
\end{align}
We next define $\mathbf{P}\mathbf{A}\triangleq\mathbf{Z}$ and $\mathbf{S}^{T}\mathbf{S}\triangleq\mathbf{R}_{S}$. Using matrix definitions in Table \ref{matrix_definitions}, we expand \eqref{interme-appendix6} as
\begin{align}\label{interme-appendix7}
\sum_{i,j}\frac{1+\hat{\theta}_{ij,\infty}}{\hat{\theta}_{ij,\infty}} &=\textrm{tr}\left(\mathbf{Z}^{-1}\mathbf{R}_{S}\mathbf{Z}\mathbf{R}_{S}\right) \notag \\&= K_{\textrm{tot}}+\underbrace{\sum_{p,q}\sum_{r,s}}_{p>r,q>s}\left(\frac{\alpha_{rs}P_{pq}}{\alpha_{pq}P_{rs}} +\frac{\alpha_{pq}P_{rs}}{\alpha_{rs}P_{pq}}\right)\rho_{pqrs}^2,\notag \\
&\geq K_{\textrm{tot}}+\underbrace{\sum_{p,q}\sum_{r,s}}_{p>r,q>s}2\rho_{pqrs}^2= \textrm{tr}\left(\mathbf{R}_{S}\mathbf{R}_{S}\right).
\end{align}
Since $\mathbf{R}_{S}$ is a symmetric matrix, its eigen decomposition is given by $\mathbf{Q}\mathbf{\Lambda}\mathbf{Q}^{T}$, where $\mathbf{Q}$ is a unitary matrix and $\mathbf{\Lambda}$ is a $K_{\textrm{tot}}\times{}K_{\textrm{tot}}$ diagonal matrix.  We note that the first $\tau$ elements in the main diagonal of $\mathbf{\Lambda}$ are the eigenvalues of $\mathbf{R}_{S}$ and the rest are zero. Accordingly, we obtain the following
\begin{align}\label{balalabah}
\textrm{tr}\left(\mathbf{R}_{S}\mathbf{R}_{S}\right)=\textrm{tr} \left(\mathbf{\Lambda}_{i}^2\right)= \sum_{i=1}^{K_{\textrm{tot}}}\lambda_{i}^{2}=\frac{\left(\sum_{i=1}^{\tau}\lambda_{i}\right)^{2}}{\tau} =\frac{K_{\textrm{tot}}^{2}}{\tau}.
\end{align}
Substituting \eqref{balalabah} into \eqref{interme-appendix7}, we obtain
\begin{align}\label{interme-appendix8}
\sum_{i,j}\frac{1+\hat{\theta}_{ij,\infty}}{\hat{\theta}_{ij,\infty}}&\geq \frac{1}{\tau}K_{\textrm{tot}}^{2}.
\end{align}

An important requirement in the multi-cell multiuser massive MIMO network is that the SINR requirement of a user cannot exceed its maximum achievable SINR with infinite $M$, i.e., $\gamma_{ij}\leq\hat{\theta}_{ij,\infty}\leq \theta_{ij,\infty}$. As such, we find that
\begin{align}\label{interme-appendix9}
\sum_{i,j}\frac{1+\hat{\theta}_{ij,\infty}}{\hat{\theta}_{ij,\infty}}\leq \sum_{i,j}\frac{1+\gamma_{ij}}{\gamma_{ij}}
\end{align}
always holds true. Using the inequality in \eqref{interme-appendix9} together with \eqref{interme-appendix8}, we obtain the desired result in \eqref{K_tot}.

\section{Proof of Lemma \ref{lemma_wbe}}\label{lemma_wbe_proof}

In this appendix, we derive the bound on the per-cell user capacity region of the WBE design. Substituting $\alpha_{ij}=\alpha$ into \eqref{assumption_pow}, we obtain the asymptotic achievable SINR of the WBE design as
\begin{align}\label{interme-appendix12}
\hat{\theta}_{lk,\infty}=\frac{P_{lk}}{\sum_{m=1}^{L}\sum_{n=1}^{K}{\rho_{lkmn}^{2}P_{mn}}-P_{lk}},
\end{align}
Using $P_{lk}=\frac{\alpha\gamma_{lk}}{1+\gamma_{lk}}$ and $\hat{\theta}_{ij,\infty}\geq\gamma_{ij}$, we rewrite \eqref{interme-appendix12} as
\begin{align}\label{interme-appendix13}
\hat{\theta}_{lk,\infty}
%&=\frac{\frac{\alpha\gamma_{lk}}{1+\gamma_{lk}}}
%{\alpha\sum_{m=1}^{L}\sum_{n=1}^{K}\frac{\rho_{lkmn}^{2}\gamma_{mn}}{1+\gamma_{mn}}
%-\frac{\alpha\gamma_{lk}}{1+\gamma_{lk}}}\notag\\
=\frac{\frac{\gamma_{lk}}{1+\gamma_{lk}}}{\sum_{p,q}\frac{\gamma_{pq}}{1+\gamma_{pq}} +\sum_{r,s}\frac{\rho_{lkrs}^{2}\gamma_{rs}}{1+\gamma_{rs}}
-\frac{\gamma_{lk}}{1+\gamma_{lk}}}\geq\gamma_{lk},
\end{align}
where $\gamma_{pq}$ is the SINR requirement of $\textrm{U}_{pq}$ using the pilot sequence $\mathbf{s}_{pq}\mid\textrm{U}_{pq}\in{}\mathcal{U}_{\mathbf{s}_{pq}}^{\textrm{WBE}}$ and $\gamma_{r,s}$ is the SINR requirement of $\textrm{U}_{rs}$ using the pilot sequence  $\mathbf{s}_{rs}\mid\textrm{U}_{rs}\in\bar{\mathcal{U}}_{\mathbf{s}_{pq}}^{\textrm{WBE}}$. Simplifying \eqref{interme-appendix13} gives
\begin{align}\label{interme-appendix14}
\sum_{p,q}\frac{\gamma_{pq}}{1+\gamma_{pq}}
+\frac{1}{\kappa}\sum_{r,s}\frac{\gamma_{rs}}{1+\gamma_{rs}}\leq1.
\end{align}
Accordingly, the bound on the per-cell user capacity region is given by
\begin{align}\label{interme-appendix149}
\sum_{s=1}^{K}\frac{\gamma_{ps}}{1+\gamma_{ps}}\leq\frac{\kappa}{L}+
\frac{\left(1-\kappa\right)\gamma_{pq}}{1+\gamma_{pq}}.
\end{align}

Jointly considering \eqref{interme-appendix149} and \eqref{BW}, we obtain the per-cell user capacity region of the WBE design, as given by \eqref{WBE}, which completes the proof.

\begin{IEEEbiography}[{\includegraphics[width=1in,height=1.25in,clip,keepaspectratio]{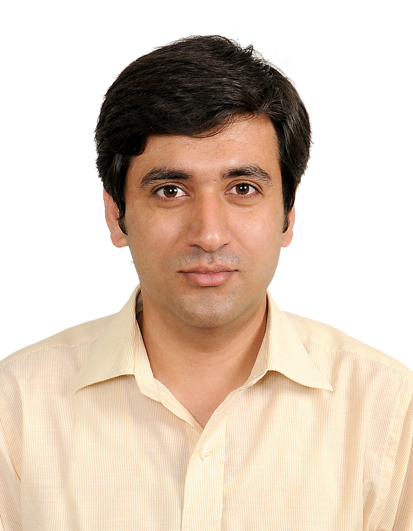}}]{Noman Akbar}(M'15)
Noman Akbar received the B.E. degree in Electrical (Telecommunication) Engineering from National University of Sciences and Technology (NUST), Rawalpindi, Pakistan, in August 2012. From August 2012 to March 2013, he worked as a researcher at NUST School of Electrical Engineering and Computer Science (SEECS). From March 2013 to February 2015, he worked in Haptics Lab, Kyung Hee University, Suwon, South Korea as a research assistant. In February 2015, he received the M.S. Degree in Computer Engineering from Kyung Hee University, Suwon, South Korea.

He is currently pursuing a PhD degree at the Research School of Engineering, Australian National University, Canberra, Australia. His research interests include wireless communications with specific focus on massive multi-antenna systems.
\end{IEEEbiography}

\begin{IEEEbiography}[{\includegraphics[width=1in,height=1.25in,clip,keepaspectratio]{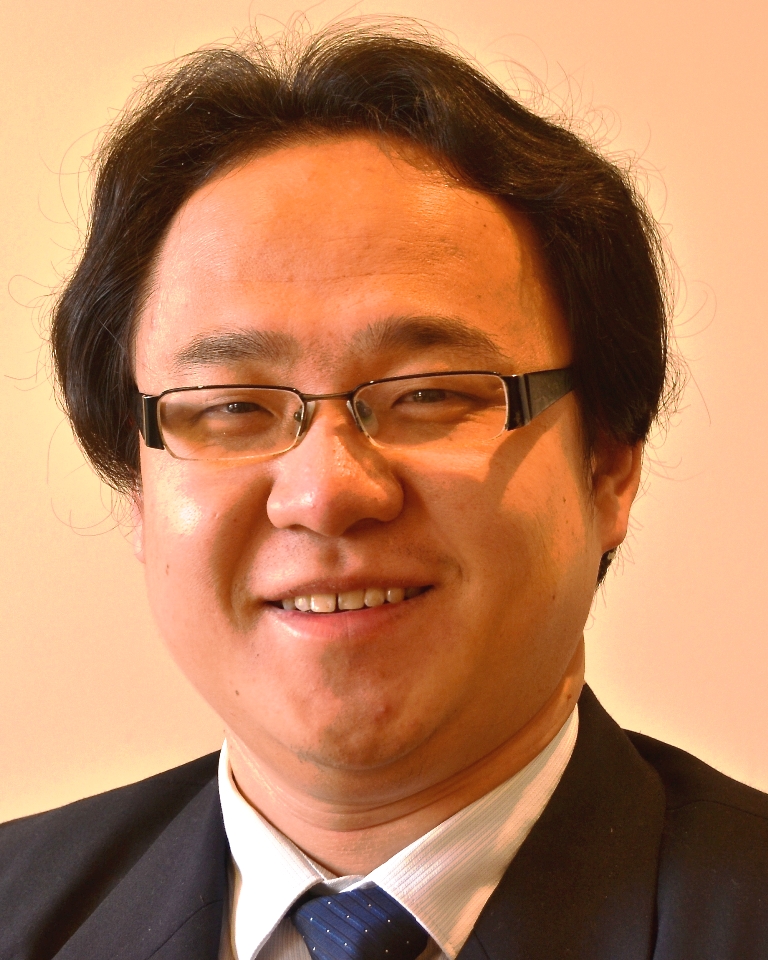}}]
{Nan Yang} (S'09--M'11) received the B.S. degree in electronics from China Agricultural University in 2005, and the M.S. and Ph.D. degrees in electronic engineering from the Beijing Institute of Technology in 2007 and 2011, respectively. He is currently a Future Engineering Research Leadership Fellow and Lecturer in the Research School of Engineering at the Australian National University. Prior to this, he was a Postdoctoral Research Fellow at the University of New South Wales (UNSW) (2012-2014) and a Postdoctoral Research Fellow at the Commonwealth Scientific and Industrial Research Organization (2010-2012). He received the Exemplary Reviewer Award of the IEEE Transactions on Communications and the Top Reviewer Award from the IEEE Transactions on Vehicular Technology in 2015, the IEEE ComSoc Asia-Pacific Outstanding Young Researcher Award and the Exemplary Reviewer Award of the IEEE Wireless Communications Letters in 2014, the Exemplary Reviewer Award of the IEEE Communications Letters in 2013 and 2012, and the Best Paper Award at the IEEE 77th Vehicular Technology Conference in 2013. He is currently serving on the Editorial Board of the IEEE Transactions on Vehicular Technology and the Transactions on Emerging Telecommunications Technologies. His general research interests lie in the areas of communications theory and signal processing, with specific interests in heterogeneous networks, massive multi-antenna systems, millimeter wave communications, cyber-physical security, and molecular communications.
\end{IEEEbiography}

\begin{IEEEbiography}[{\includegraphics[width=1in,height=1.25in,clip,keepaspectratio]{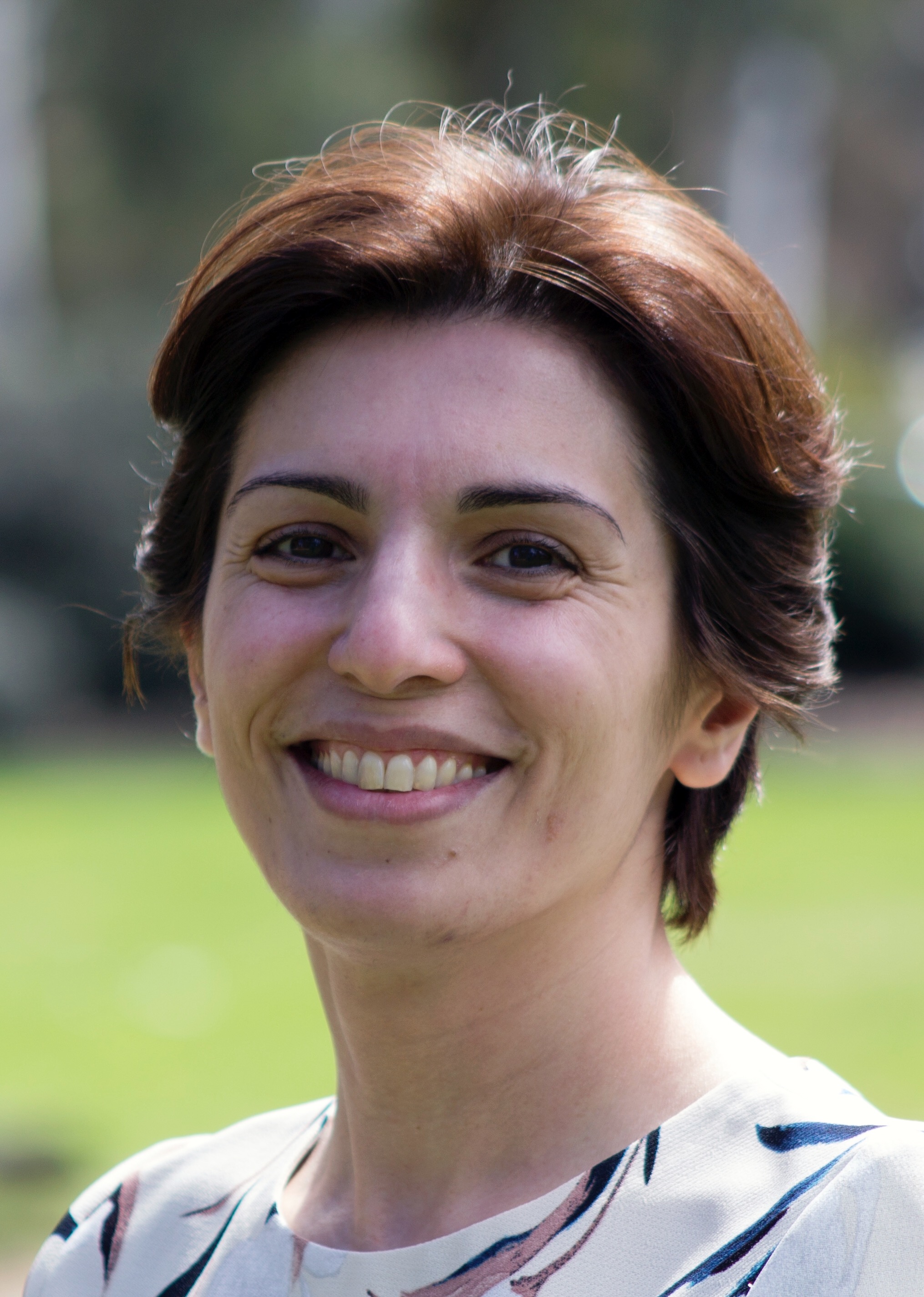}}]
{Parastoo Sadeghi} (S'02--M'06--SM'07) is an Associate Professor at the Research School of Engineering, Australian National University, Canberra, Australia. She received the BSc and MSc degrees in electrical engineering from Sharif University of Technology, Tehran, Iran, in 1995 and 1997, respectively, and the PhD degree in electrical engineering from the University of New South Wales, Sydney, Australia, in 2006. From 1997 to 2002, she was a Research Engineer and then a Senior Research Engineer at Iran Communication Industries, Tehran, and at Deqx (formerly known as Clarity Eq), Sydney. She has visited various research institutes, including the Institute for Communications Engineering, Technical University of Munich, in 2008 and MIT in 2009 and 2013.

Dr. Sadeghi has co-authored more than 140 refereed journal or conference papers and a book on Hilbert Space Methods in Signal Processing (Cambridge Univ. Press, 2013). She is currently serving as an Associate Editor of the IEEE Transactions on Information Theory. Dr. Sadeghi has been a Chief Investigator in a number of Australian Research Council Discovery and Linkage Projects. Her research interests are mainly in the areas of network coding, information theory, wireless communications theory, and signal processing.
\end{IEEEbiography}

\begin{IEEEbiography}[{\includegraphics[width=1in,height=1.25in,clip,keepaspectratio]{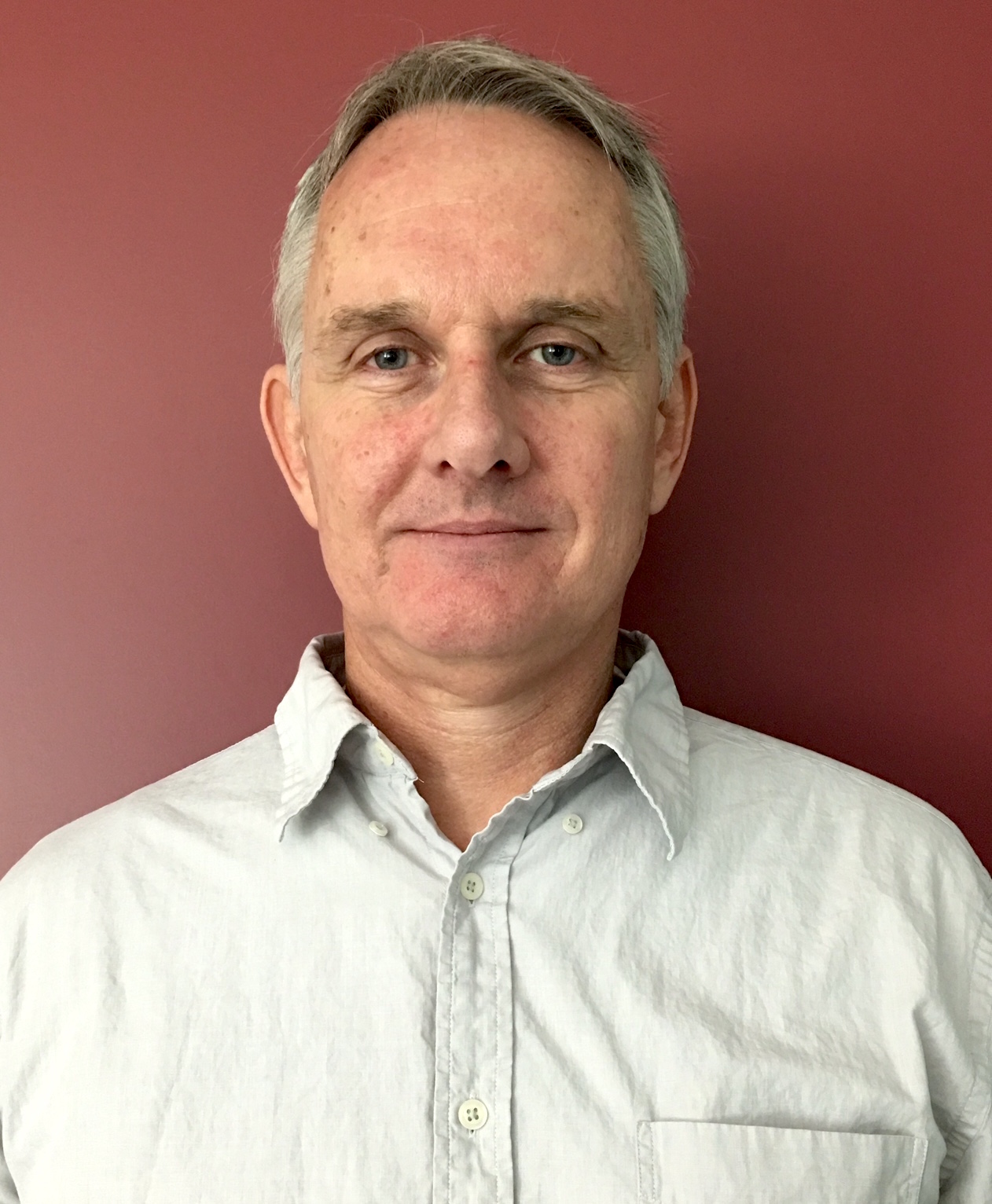}}]
{Rodney A. Kennedy}(S'86--M'88--SM'01--F'05) received the B.E. degree (1st class honours and university medal) from the University of New South Wales, Sydney, Australia, the M.E. degree from the University of Newcastle, and the Ph.D. degree from the Australian National University, Canberra. Since 2000 he has been is a Professor in engineering at the Australian National University, Canberra, Australia.
He has co-authored close to 400 refereed journal or conference papers and a book “Hilbert Space Methods in Signal Processing” (Cambridge Univ. Press, 2013). He has been a Chief Investigator in a number of Australian Research Council Discovery and Linkage Projects. His research interests include digital signal processing, digital and wireless communications, and acoustical signal processing.

\end{IEEEbiography}

\end{document}